\documentclass[12pt]{article}
\usepackage{graphicx}
\def \d {{\rm d}}

\begin{document}

\title{\bf A new look at the Pleba\'nski--Demia\'nski family of solutions}

\author{J. B. Griffiths$^1$ and J. Podolsk\'y$^2$
\\ \\ \small
$^1$Department of Mathematical Sciences, Loughborough University, \\
\small Loughborough,  Leics. LE11 3TU, U.K. \\ 
\small $^2$Institute of Theoretical Physics, Charles University in Prague,\\
\small V Hole\v{s}ovi\v{c}k\'ach 2, 18000 Prague 8, Czech Republic.}

\date{\today}
\maketitle

\begin{abstract}
\noindent
The Pleba\'nski--Demia\'nski metric, and those that can be obtained from it by taking coordinate transformations in certain limits, include the complete family of space-times of type~D with an aligned electromagnetic field and a possibly non-zero cosmological constant. Starting with a new form of the line element which is better suited both for physical interpretation and for identifying different subfamilies, we review this entire family of solutions. Our metric for the expanding case explicitly includes two parameters which represent the acceleration of the sources and the twist of the repeated principal null congruences, the twist being directly related to both the angular velocity of the sources and their NUT-like properties. The non-expanding type~D solutions are also identified. All special cases are derived in a simple and transparent way. 
\end{abstract}

\newpage
\tableofcontents

\newpage
\section{Introduction}

The complete family of type D space-times with an aligned non-null electromagnetic field and a possibly non-zero cosmological constant $\Lambda$ can be represented by a metric that was given by Pleba\'nski and Demia\'nski \cite{PleDem76} together with those that can be derived from it by certain transformations and limiting procedures. 
These solutions are characterised by two related quartic functions whose coefficients are determined by seven arbitrary parameters which include $\Lambda$ and both electric and magnetic charges. 
For the vacuum case with vanishing cosmological constant, they include all the particular solutions that were identified by Kinnersley \cite{Kinnersley69}. For the sub-cases in which the repeated principal null congruences are expanding, these metrics have been analysed further by Weir and Kerr \cite{WeirKerr77}, where the relations between the different forms of the line element can be deduced. (They have also been given independently by Debever and Kamran \cite{DebKam80} and Ishikawa and Miyashita \cite{IshMiy82}.)

Unfortunately, many particular and well-known type D space-times are not included explicitly in the original form of the line element. They can only be obtained from it by using certain degenerate transformations. Moreover, the parameters that were introduced in the original papers are not the most useful ones for a physical interpretation of the solutions. One purpose of this paper is therefore to present a new form of the Pleba\'nski--Demia\'nski metric in which the parameters are given a clear physical meaning and from which the various special cases can be obtained in a more satisfactory way. In this way we will clarify the complete family of solutions.

A new look at these solutions seems to us to be particularly important at the present time because they are now being used in new ways. People working in semi-classical quantum gravity have used these metrics to investigate the pair production of black holes in cosmological backgrounds (see e.g. \cite{HawRoss95}--\cite{BooMan99}). Others are working to extend these solutions to higher dimensions. Yet these solutions are still not well-understood at the classical level of general relativity. In particular, the physical significance of the parameters employed in the original forms are only properly identified in the most simplified special cases. Thus the emphasis here is not to look for the most general metric form which covers all cases. This was achieved many years ago with the work of Carter \cite{Carter68}, Debever {\it et al} \cite{DeKaMcL84}, Garc\'{\i}a~D. \cite{GarciaD84} and others. Rather, our purpose is to cast the metric in a form in which the parameters employed have clear physical interpretations and thus to classify the complete family of solutions in a way that clarifies their physical properties.

In order to achieve this, we will first modify the original Pleba\'nski--Demia\'nski metric to include two parameters, denoted by $\alpha$ and $\omega$, which respectively represent the acceleration of the sources and the twist of the repeated principal null congruences. Then we will derive a more general form of the line element which explicitly contains all the well-known special subfamilies, at least for the cases in which the repeated principal null congruence is diverging and the orbits of the Killing vectors are non-null. 
(The case in which the group orbits are null is not considered here.)

The solutions contained in this family are characterized by two generally quartic functions whose coefficients are related to the physical parameters of the space-time. One of our main purposes is to clarify the physical meaning of these coefficients which have traditionally been misinterpreted in the general case. In particular, we locate the cosmological constant in the most appropriate place and identify the relation between the Pleba\'nski--Demia\'nski parameter $n$ and the NUT parameter. It has also been traditional to use an available coordinate freedom to remove the linear term in one of these quartics. 
However, as noted by Hong and Teo in a particular case \cite{HongTeo03}, \cite{HongTeo05}, it is much more helpful to use the freedom to simplify the roots of this quartic. It will be shown that this helps very significantly both in the interpretation of these solutions and also in performing the associated calculations.

In section~3, we will review the family of solutions of this type which possess space-like surfaces of positive curvature. These include the solutions which are considered to have greatest physical significance as they relate to the fields around generalised forms of black holes. After that, we consider the complete families of non-accelerating and then non-twisting solutions in more detail.

In practice, it is convenient to distinguish the cases in which the repeated principal null directions are either expanding or non-expanding. A separate general form of the metric is required for the non-expanding case. Such a form is derived in section~\ref{nonexpsolns} using a degenerate transformation of our initial metric. As the twist also vanishes in this limit, it follows that these solutions must be type~D members of the solutions of Kundt's class -- a result that is demonstrated explicitly.

Finally, we identify the members of this family of solutions which reduce to the Bertotti--Robinson solution, or other direct product space-times, in an appropriate limit.

\section{The Pleba\'nski--Demia\'nski metric}

We consider here a general family of type~D solutions of Einstein's equations including a generally non-zero cosmological constant $\Lambda$. These may be vacuum or include a non-null electromagnetic field such that the two repeated principal null congruences of the Weyl tensor are aligned\footnote{
          Other non-null type~D electrovacuum solutions exist in which only one
of the principal null congruences of the electromagnetic field is aligned with a
repeated principal null congruences of the Weyl tensor (see for example
\cite{KowPleb77}). However, such solutions are not considered here.} 
          with the two principal null congruences of the non-null electromagnetic field.

\subsection{The original form of the metric}

Let us start with the Pleba\'nski--Demia\'nski metric \cite{PleDem76} (see \S21.1.2 of \cite{SKMHH03}) which is given by 
  \begin{equation} 
 \d s^2={1\over(1-\hat p\hat r)^2} \Bigg[
{{\cal Q}(\d\hat\tau-\hat p^2\d\hat\sigma)^2\over\hat r^2+\hat p^2} -{{\cal P}(\d\hat\tau+\hat r^2\d\hat\sigma)^2\over\hat r^2+\hat p^2} 
 -{\hat r^2+\hat p^2\over{\cal P}}\,\d\hat p^2
-{\hat r^2+\hat p^2\over{\cal Q}}\,\d\hat r^2 \Bigg] \,,
  \label{oldPDMetric}
  \end{equation}  
  where 
  \begin{equation} 
  \begin{array}{l} 
  {\cal P} =\hat k +2\hat n\hat p -\hat\epsilon\hat p^2 
  +2\hat m\hat p^3-(\hat k+\hat e^2+\hat g^2+\Lambda/3)\hat p^4 \,, \\[8pt] 
  {\cal Q} =(\hat k+\hat e^2+\hat g^2) -2\hat m\hat r +\hat\epsilon\hat r^2 
  -2\hat n\hat r^3-(\hat k+\Lambda/3)\hat r^4 \,,
  \end{array} 
 \label{oldPQeqns}
  \end{equation} 
  and $\hat m$, $\hat n$, $\hat e$, $\hat g$, $\hat\epsilon$, $\hat k$ and $\Lambda$ are arbitrary real parameters\footnote{
      Debever \cite{Debever71} had previously found a type~D metric with an equivalent set of seven arbitrary parameters. 
     }. It is often assumed that $\hat m$ and $\hat n$ are the mass and NUT parameters respectively, although this is not generally the case. The parameters $\hat e$ and $\hat g$ represent electric and magnetic charges. The parameter $\gamma$ that is used in \cite{PleDem76} and \cite{SKMHH03} is obtained by putting \ $\hat k=\gamma-\hat g^2-\Lambda/6$. \ However, as shown in \cite{PodGri01}, it is more convenient for physical interpretation to include the cosmological constant in the form given in~(\ref{oldPQeqns}).

As is required for type~D space-times of this type, the general family of solutions represented by (\ref{oldPDMetric}) admits (at least) two commuting Killing vectors $\partial_{\hat\sigma}$ and $\partial_{\hat\tau}$ whose orbits are spacelike in regions with ${\cal Q}>0$ and timelike when ${\cal Q}<0$. Type~D solutions also exist\footnote{
      Metrics in which the group orbits are null can be obtained by using a degenerate coordinate transformation. This is described in detail by Garc\'{\i}a and Pleba\'nski \cite{GarPle82} (see also p322 of \cite{SKMHH03}) who confirm that a particular case of Leroy \cite{Leroy78} is included in this general family. All such type~D space-times with null group orbits have been given by Garc\'{\i}a and Salazar \cite{GarSal83}. A generalized form of the metric (\ref{oldPDMetric}) which includes all known cases of this type has been given by Debever, Kamran and McLenaghan \cite{DeKaMcL83}, \cite{DeKaMcL84}, although this form is not well suited for the interpretation of the solutions. A form of the metric which covers the cases of null and non-null orbits simultaneously was used by Garc\'{\i}a \cite{GarciaD84}. (See also \cite{GarMac98}.)
      } in which the group orbits are null, but these are not considered here.

\subsection{A modified form of the metric}

For purposes of interpreting the Pleba\'nski--Demia\'nski metric, it is convenient (see \cite{Pleb75} and \cite{GriPod05}) to introduce the rescaling 
  \begin{equation} 
  \hat p=\sqrt{\alpha\omega}\,p, \qquad \hat r=\sqrt{\alpha\over\omega}\,r, \qquad  \hat\sigma=\sqrt{\omega\over\alpha^3}\,\sigma, \qquad 
 \hat\tau=\sqrt{\omega\over\alpha}\,\tau,
  \label{scaling}
  \end{equation} 
 with the relabelling of parameters 
  \begin{equation} 
  \hat m+i\hat n=\Big({\alpha\over\omega}\Big)^{3/2}(m+in), \qquad
  \hat e+i\hat g={\alpha\over\omega}(e+ig), \qquad
 \hat\epsilon={\alpha\over\omega}\,\epsilon, \qquad
 \hat k=\alpha^2k. 
  \label{scaleps}
  \end{equation} 
 This introduces two additional parameters $\alpha$ and $\omega$. With these changes, the metric becomes 
  \begin{equation}
  \begin{array}{r}
{\displaystyle  \d s^2={1\over(1-\alpha pr)^2} \Bigg[
{Q\over r^2+\omega^2p^2}(\d\tau-\omega p^2\d\sigma)^2
  -{P\over r^2+\omega^2p^2}(\omega\d\tau+r^2\d\sigma)^2 } \hskip3pc \\[12pt]
  {\displaystyle -{r^2+\omega^2p^2\over P}\,\d p^2
-{r^2+\omega^2p^2\over Q}\,\d r^2 \Bigg],}
  \end{array}
  \label{PleDemMetric}
  \end{equation}
  where
 \begin{equation}
 \begin{array}{l}
 P=P(p) =k +2\omega^{-1}np -\epsilon p^2 +2\alpha mp^3
-\big[\alpha^2(\omega^2 k+e^2+g^2)+\omega^2\Lambda/3\big]p^4 \\[8pt]
 Q=Q(r) =(\omega^2k+e^2+g^2) -2mr +\epsilon r^2 -2\alpha\omega^{-1}nr^3
-(\alpha^2k+\Lambda/3)r^4,
 \end{array}
 \label{PQeqns}
 \end{equation} 
 and $m$, $n$, $e$, $g$, $\Lambda$, $\epsilon$, $k$, $\alpha$ and $\omega$ 
are arbitrary real parameters of which two can be chosen for convenience. It should be emphasised that, apart from $\Lambda$, $e$ and $g$, the parameters included in this metric do not necessarily have their traditional physical interpretation. They only acquire their usual specific well-identified meanings in certain special sub-cases.

Adopting the null tetrad 
 \begin{equation}
 \begin{array}{l}
 l^\mu ={\displaystyle {(1-\alpha pr)\over\sqrt{2(r^2+\omega^2p^2)}}\left[ 
 {1\over\sqrt Q}\Big(r^2\partial_\tau-\omega\partial_\sigma\Big) 
-\sqrt Q\,\partial_r \right]} \,, \\[16pt] 
 n^\mu ={\displaystyle {(1-\alpha pr)\over\sqrt{2(r^2+\omega^2p^2)}}\left[ 
 {1\over\sqrt Q}\Big(r^2\partial_\tau-\omega\partial_\sigma\Big) 
+\sqrt Q\,\partial_r \right]} \,, \\[16pt] 
 m^\mu ={\displaystyle {(1-\alpha pr)\over\sqrt{2(r^2+\omega^2p^2)}}\left[ 
 -{1\over\sqrt P}\Big(\omega p^2\partial_\tau+\partial_\sigma\Big) 
+i\sqrt P\,\partial_p \right]} \,,  
 \end{array} 
 \label{GeneralTetrad}
 \end{equation} 
 the spin coefficients are given by 
 \begin{eqnarray}
 &&\kappa = \sigma = \lambda = \nu = 0 \ ,\nonumber\\[6pt] 
 &&\rho = \mu =\sqrt{\frac{Q}{2(r^2+\omega^2p^2)}}\,
{(1+i\alpha\omega p^2)\over(r+i\omega p)}\  ,\nonumber\\[6pt] 
 &&\tau = \pi = \sqrt{\frac{P}{2(r^2+\omega^2p^2)}}\,
{(\omega-i\alpha r^2)\over(r+i\omega p)}\ ,\label{spincoeffts}\\[6pt] 
 &&\epsilon = \gamma = {1\over4}\sqrt{\frac{Q}{2(r^2+\omega^2p^2)}}
\left[ 2{(1-\alpha pr)\over(r+i\omega p)} -2\alpha p 
-(1-\alpha pr){Q'\over Q} \right] \ ,\nonumber\\[6pt] 
 &&\alpha = \beta ={1\over4}\sqrt{P\over2(r^2+\omega^2p^2)} 
\left[ 2\omega{(1-\alpha pr)\over(r+i\omega p)} +2i\alpha r 
+i(1-\alpha pr){P'\over P} \right] \ .\nonumber 
 \end{eqnarray} 
 These indicate that the congruences tangent to $l^\mu$ and $n^\mu$ are both
geodesic and shear-free but have non-zero expansion. It can be seen that the
twist of both congruences is proportional to $\omega$. In some particular cases $\omega$ is directly related to both the angular velocity of sources and the effects of the NUT parameter (see \cite{GriPod05}).

In terms of the tetrad (\ref{GeneralTetrad}), the only non-zero component of the
Weyl tensor is 
 \begin{equation} 
 \Psi_2=-(m+in)\left({1-\alpha pr\over r+i\omega p}\right)^3 
+(e^2+g^2)\left({1-\alpha pr\over r+i\omega p}\right)^3 
{1+\alpha pr\over r-i\omega p}. 
 \label{Weyl1} 
 \end{equation}  
This confirms that these space-times are of algebraic type~D, and that the tetrad vectors $l^\mu$ and $n^\mu$ as chosen above are aligned with the repeated
principal null directions of the Weyl tensor. The only non-zero components of
the Ricci tensor are 
 \begin{equation} 
 \Phi_{11}= {1\over2}\,(e^2+g^2)\,{(1-\alpha pr)^4\over(r^2+\omega^2p^2)^2}, 
 \qquad \Lambda_{\rm NP}={\textstyle{1\over6}}\,\Lambda. 
 \label{Ricci1} 
 \end{equation}  
 Together, these indicate the presence of a curvature singularity at $p=r=0$ which, if contained within the space-time, may be considered as the source of the gravitational field.

Having introduced $\alpha$ and $\omega$ as continuous parameters, we are free to
use the rescaling (\ref{scaling}) with (\ref{scaleps}) to scale the parameters $\epsilon$ and $k$ to some specific values (without changing their signs). For example, we could set them to the values $+1$, $0$ or $-1$, but it will generally be more convenient to scale them to some other appropriate values. It is clear that $e$ and $g$ are the electric and magnetic charges of the sources and $\Lambda$ is the cosmological constant. For certain choices of the other parameters, it is found that $m$ is related to the mass of the source and $n$ is related to the NUT parameter (although it should not be identified with it in general).

To retain a Lorentzian signature in (\ref{PleDemMetric}), it is necessary that
$P>0$. And, since $P(p)$ is generally a quartic function, the coordinate $p$ must be restricted to a particular range between appropriate roots. If this range includes $p=0$, it is necessary that $k>0$ and, in this case, it is always possible to use the scaling (\ref{scaling}) to set $k=1$. However, this is not necessary, and we simply note that the requirement that $P>0$ may place some restriction on the possible signs of the parameters $\epsilon$ and~$k$. In fact, various particular cases can be identified and classified according to the number and types of the roots of $P$ and the range of $p$ that is adopted (see e.g. \cite{IshMiy82}). Points at which $P=0$ generally correspond to poles of the coordinates. By contract, surfaces on which $Q=0$ are horizons through which coordinates can be extended. It is also significant that 
 $$ Q(r)= -\alpha^2r^4P({1\over\alpha r}) -{\Lambda\over3}\left({\omega^2\over\alpha^2}+r^4\right). $$ 
 When $\Lambda=0$, this relates the number and character of the roots of these two quartics. However, when $\Lambda\ne0$, this correspondence is obscured.

The line element (\ref{PleDemMetric}) is flat if $m=n=0$, $e=g=0$ and
$\Lambda=0$, but the remaining parameters $\epsilon$, $k$, $\alpha$ and
$\omega$ may be non-zero in this flat limit. Moreover, it is not immediately obvious that the metric (\ref{PleDemMetric}) includes the
Schwarzschild--de~Sitter solution, the Reissner--Nordstr\"om solution, the Kerr metric, the NUT solution or the $C$-metric, which are all known to be of algebraic type~D, or the Robinson--Trautman type~D space-times. In the following sections, it will be shown that a simple transformations of (\ref{PleDemMetric}) leads to a form which explicitly includes all these well-known special cases.

\section{A general metric for expanding solutions}
\label{expanding}

When $\Lambda=0$, the line element (\ref{PleDemMetric}) already contains the Kerr--Newman solution for a charged rotating black hole. It also contains the charged $C$-metric for accelerating black holes. However, it does not include the type~D non-singular NUT solution \cite{NewTamUnt63}. To cover all these cases and their generalizations, it is necessary to introduce a specific shift in the coordinate~$p$. In fact, this procedure is essential to obtain the correct metric for accelerating and rotating black holes. We therefore start with the metric (\ref{PleDemMetric}) with (\ref{PQeqns}), and perform the coordinate transformation 
  \begin{equation}
  p={l\over\omega}+{a\over\omega}\,\tilde p, \qquad  
  \tau=t-{(l+a)^2\over a}\,\phi, 
  \qquad \sigma=-{\omega\over a}\,\phi,
  \label{trans1A}
  \end{equation}
  where $a$ and $l$ are new arbitrary parameters. By this procedure, we obtain the metric  
  \begin{equation}
  \begin{array}{l}
{\displaystyle \d s^2={1\over\Omega^2}\left\{
{Q\over\rho^2}\left[\d t- \left(a(1-\tilde p^2)
+2l(1-\tilde p) \right)\d\phi \right]^2
   -{\rho^2\over Q}\,\d r^2 \right.
} \\[8pt]
  \hskip8pc {\displaystyle 
 \left. -{\tilde P\over\rho^2} \Big[ a\d t  
  -\Big(r^2+(l+a)^2\Big)\d\phi \Big]^2  
-{\rho^2\over\tilde P}\,\d\tilde p^2 \right\}, }
\end{array}
  \label{altPlebMetric}
  \end{equation}
  where
  $$ \begin{array}{l}
  \Omega=1-{\displaystyle{\alpha\over\omega}} (l+a\tilde p)r \,, \\[6pt]
  \rho^2 =r^2+(l+a\tilde p)^2 \,, \\[6pt]
  \tilde P= a_0 +a_1\tilde p +a_2\tilde p^2
+ a_3\tilde p^3 +a_4\tilde p^4 \,, \\
  Q= {\displaystyle (\omega^2k+e^2+g^2) -2mr +\epsilon r^2 
-2\alpha{n\over\omega} r^3 -\left(\alpha^2k+{\Lambda\over3}\right)r^4},
  \end{array} $$   
  and we have put
  $$ \begin{array}{l}
  a_0={\displaystyle {1\over a^2} \left( \omega^2k +2nl -\epsilon l^2
+2\alpha{l^3\over\omega} m 
-\bigg[{\alpha^2\over\omega^2}(\omega^2k+e^2+g^2)+{\Lambda\over3}\bigg]l^4 \right) }\,, \\[6pt]
  a_1={\displaystyle {2\over a}\left( n -\epsilon l +3\alpha{l^2\over\omega}m 
-2\bigg[{\alpha^2\over\omega^2}(\omega^2k+e^2+g^2)+{\Lambda\over3}\bigg]l^3 \right) }\,, \\[6pt]
  a_2={\displaystyle -\epsilon +6\alpha{l\over\omega}m -6\bigg[{\alpha^2\over\omega^2}(\omega^2k+e^2+g^2)+{\Lambda\over3}\bigg]l^2 }\,, \\[6pt]
  a_3={\displaystyle 2\alpha{a\over\omega}m 
  -4\bigg[{\alpha^2\over\omega^2}(\omega^2k+e^2+g^2)+{\Lambda\over3}\bigg]al }\,, \\[6pt]
  a_4={\displaystyle -\bigg[{\alpha^2\over\omega^2}(\omega^2k+e^2+g^2)+{\Lambda\over3}\bigg]a^2 }\,.
  \end{array} $$

These solutions generally have seven essential parameters $m$, $n$, $e$, $g$, $\alpha$, $\omega$ and $\Lambda$. They also have two parameters $k$ and $\epsilon$ which can be scaled to any convenient values. In addition, we have the further parameters $a$ and $l$ which can be chosen arbitrarily. In practice, it is convenient to choose $a$ and $l$ to satisfy certain conditions which simplify the form of the metric, and then to re-express $n$ and $\omega$ in terms of these parameters.

The properties of the solutions in this family depend critically on the character of the function $\tilde P(\tilde p)$. In fact, as an arbitrary quartic, $\tilde P$ can have up to four distinct roots, and Lorentzian space-times only occur for ranges of $\tilde p$ for which $\tilde P>0$. When more than one such range exists, the different possibilities correspond to distinct space-times which have different physical interpretations.

When $\tilde P$ has no roots, this function can only be positive and $\tilde p\in(-\infty,\infty)$.

For the cases in which $\tilde P$ has at least one root, without loss of generality we can choose the parameters $a$ and $l$ so that such a root occurs at $\tilde p=1$. The metric (\ref{altPlebMetric}) is then regular at $\tilde p=1$ which corresponds to a coordinate pole on an axis, and it is then appropriate to take $\phi$ as a periodic coordinate.

When another distinct root of $\tilde P$ exists, it is always possible to exhaust the freedom in $a$ and $l$ to set the second root at $\tilde p=-1$. The metric component $a(1-\tilde p^2)$ is then regular at this second pole while the component $2l(1-\tilde p)$ is not. Thus, the metric is regular at $\tilde p=1$, but a singularity of some kind occurs at $\tilde p=-1$. (In fact, unless $l=0$, the region near $\tilde p=-1$ contains closed timelike lines.) With this choice, and for the positive curvature case in which both poles are located on a continuous axis, it will be shown that $a$ corresponds to a Kerr-like rotation parameter for which the corresponding metric components are regular on the entire axis, while $l$ corresponds to a NUT parameter for which the corresponding components are only regular on the half-axis $\tilde p=1$.

We have now introduced through (\ref{trans1A}) a shift and scaling of $p$ such that, if $\tilde P$ has at least two roots, then it admits the two factors $(1-\tilde p)$ and $(1+\tilde p)$. Thus 
 $$ \tilde P= (1-\tilde p^2)(a_0-a_3\tilde p-a_4\tilde p^2), $$ 
 which implies that the above coefficients must satisfy the conditions  
 \begin{equation}
 a_1+a_3=0, \qquad a_0+a_2+a_4=0. 
 \label{Pconditions}  
 \end{equation}
 These conditions provide two linear equations which specify the two parameters $\epsilon$ and $n$ in terms of $a$ and $l$ as 
 \begin{eqnarray}
 &&\epsilon= {\omega^2k\over a^2-l^2}+4\alpha{l\over\omega}\,m 
 -(a^2+3l^2) \left[ {\alpha^2\over\omega^2}(\omega^2k+e^2+g^2)+{\Lambda\over3} \right], \label{epsilon}\\[8pt]
 &&n= {\omega^2k\,l\over a^2-l^2} -\alpha{(a^2-l^2)\over\omega}\,m 
 +(a^2-l^2)l \left[ {\alpha^2\over\omega^2}\,(\omega^2k+e^2+g^2)+{\Lambda\over3} \right].
  \label{n}  
 \end{eqnarray}
 Equation (\ref{n}) explicitly relates the Pleba\'nski--Demia\'nski parameter $n$ to the NUT parameter $l$. With these definitions, we then obtain that 
 $$ a_0={\omega^2k\over a^2-l^2} -2\alpha{l\over\omega}\,m 
 +3\alpha^2{l^2\over\omega^2}(\omega^2k+e^2+g^2) +l^2\Lambda. $$

The character of the solution then partly depends on whether this expression for $a_0$ is positive, negative or zero. If it is non-zero, the scaling freedom can then be used to set $a_0=\pm1$. This equation then effectively defines the parameter $k$. Thus, for any given value of $a_0$, the constant $k$ is given by 
  \begin{equation}
  \left( {\omega^2\over a^2-l^2}+3\alpha^2l^2 \right)\,k 
  =a_0 +2\alpha{l\over\omega}\,m 
  -3\alpha^2{l^2\over\omega^2}(e^2+g^2) -l^2\Lambda. 
  \label{k}
  \end{equation}

The original metric (\ref{oldPDMetric}) contained the three parameters $\hat n$, $\hat\epsilon$ and $\hat k$. In the above argument, we have increased the number of such parameters to include $n$, $\epsilon$, $k$, $\alpha$, $\omega$, $a$ and $l$, but we have also introduced three constraints that are effectively represented by (\ref{epsilon}), (\ref{n}) and (\ref{k}). One remaining (scaling) freedom is therefore still available which may be used to set $\omega$ to any convenient value (assuming $a$ and $l$ do not both vanish). There thus remain the three parameters $\alpha$, $a$ and $l$ in addition to $m$, $e$, $g$ and $\Lambda$.

\section{Generalised black holes ($\tilde P$ has two roots, $a_0=1$)}
\label{poscurv}

In this section, we will concentrate on the physically most relevant particular case of the line element (\ref{altPlebMetric}) for which $\tilde P$ has at least two distinct roots and $a_0>0$, so that we can set $a_0=1$. In this case, the surfaces spanned by $\tilde p$ and $\phi$ have positive curvature.

\subsection{The complete family of black hole-like space-times}

For the case considered here, $\tilde p$ is taken to cover the range between the roots $\tilde p=\pm1$ and it is natural to put $\tilde p=\cos\theta$, where $\theta\in[0,\pi]$. In this case, the metric (\ref{altPlebMetric}) becomes 
  \begin{equation}
  \begin{array}{l}
{\displaystyle \d s^2={1\over\Omega^2}\left\{
{Q\over\rho^2}\left[\d t- \left(a\sin^2\theta
+4l\sin^2{\textstyle{\theta\over2}} \right)\d\phi \right]^2
   -{\rho^2\over Q}\,\d r^2 \right.
} \\[8pt]
  \hskip8pc {\displaystyle 
 \left. -{\tilde P\over\rho^2} \Big[ a\d t  
  -\Big(r^2+(a+l)^2\Big)\d\phi \Big]^2  
-{\rho^2\over\tilde P}\sin^2\theta\,\d\theta^2 \right\}, }
\end{array}
  \label{newMetric}
  \end{equation}
  where 
  \begin{equation}
  \begin{array}{l}
  {\displaystyle \Omega=1-{\alpha\over\omega}(l+a\cos\theta)\,r } \\[6pt]
  \rho^2 =r^2+(l+a\cos\theta)^2 \\[6pt]
  \tilde P= \sin^2\theta\,(1-a_3\cos\theta-a_4\cos^2\theta) \\
  Q= {\displaystyle (\omega^2k+e^2+g^2) -2mr +\epsilon r^2 
-2\alpha{n\over\omega} r^3 -\left(\alpha^2k+{\Lambda\over3}\right)r^4} 
  \end{array} 
  \label{newMetricFns}
  \end{equation} 
  and 
 \begin{equation}
 \begin{array}{l}
  {\displaystyle a_3= 2\alpha{a\over\omega}m -4\alpha^2{al\over\omega^2}
  (\omega^2k+e^2+g^2) -4{\Lambda\over3}al } \\[6pt]
  {\displaystyle a_4= -\alpha^2{a^2\over\omega^2}(\omega^2k+e^2+g^2) 
  -{\Lambda\over3}a^2 }
  \end{array}
 \label{a34}
 \end{equation} 
 with $\epsilon$, $n$ and $k$ given by (\ref{epsilon})--(\ref{k}). It is also assumed that $|a_3|$ and $|a_4|$ are sufficiently small that $\tilde P$ has no additional roots with $\theta\in[0,\pi]$. 
This solution contains eight arbitrary parameters $m$, $e$, $g$, $a$, $l$, $\alpha$, $\Lambda$ and $\omega$. Of these, the first seven can be varied independently, and $\omega$ can be set to any convenient value if $a$ or $l$ are not both zero.

\begin{figure}[hbt]
\begin{center} \includegraphics[scale=0.85, trim=5 5 5 -5]{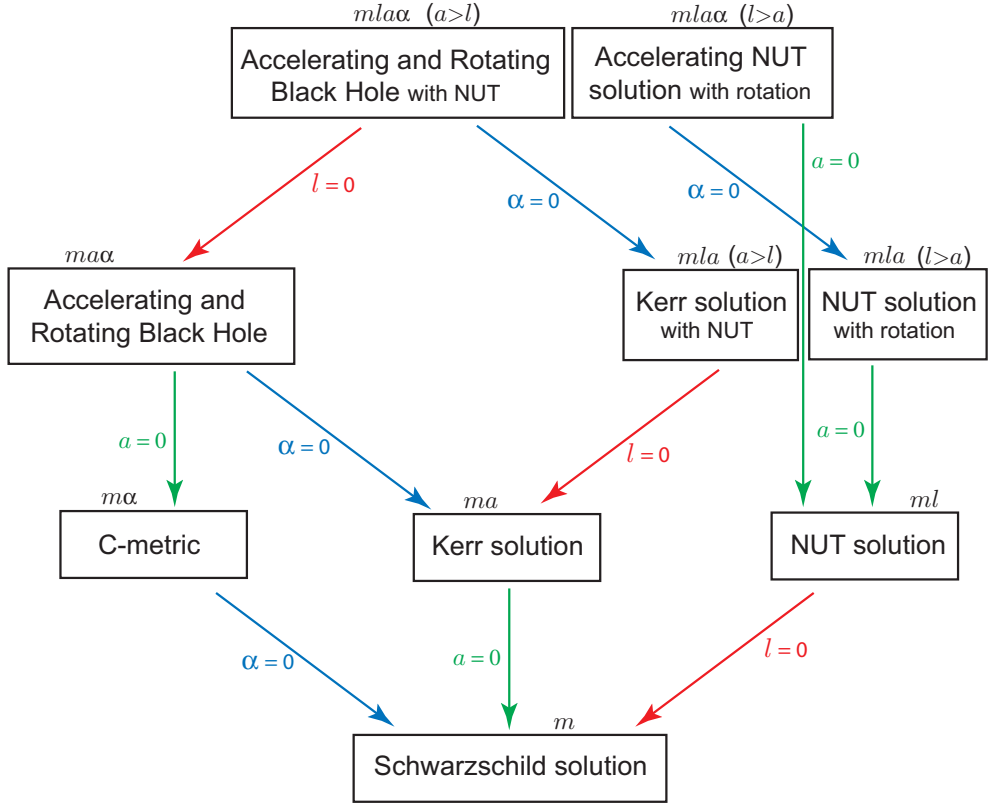} 
\caption{ \small The structure of the family of solutions represented by (\ref{newMetric}) when $\Lambda=0$, $e=g=0$ and \hbox{$m\ne0$}. This family has four parameters $m$, $l$, $a$ and $\alpha$. An accelerating Kerr solution with a small NUT parameter has been distinguished from an accelerating NUT solution with a small rotation as their singularity structures differ significantly even though their metric forms are identical. For the same reason, the Kerr--NUT solution has similarly been divided. An accelerating NUT solution without rotation has not been identified. All the special cases indicated have obvious charged versions and versions with a non-zero cosmological constant. }
\label{AccRotBHstructure}
\end{center}
\end{figure}

It was shown in \cite{GriPod05} that, when $\Lambda=0$, the metric (\ref{newMetric}) represents an accelerating and rotating charged black hole with a generally non-zero NUT parameter. However, an arbitrary cosmological constant is now included so that the background is either Minkowski, de~Sitter or anti-de~Sitter space-time. For the vacuum case in which $\Lambda=0$ and $e=g=0$, the general structure of this family of solutions is given in figure~\ref{AccRotBHstructure}. The special cases are generally well-known and will not be discussed here.

The non-zero components of the curvature tensor are given by (\ref{Weyl1}) and (\ref{Ricci1}) in which $\omega p$ is replaced by \ $l+a\cos\theta$. \ It can be seen that the metric (\ref{newMetric}) has a curvature singularity when \ $\rho^2=0$. \ If \ $|l|\le|a|$, \ this occurs when \ $r=0$ \ and \ $\cos\theta=-l/a$. \ On the other hand, if \ $|l|>|a|$, \ $\rho^2$ cannot be zero and the metric is non-singular. These two cases have to be considered separately as they clearly have very different global and singularity structures. This distinction is also indicated in figure~\ref{AccRotBHstructure}.

\goodbreak
When $a^2\ge l^2$, the metric (\ref{newMetric}) has a curvature singularity at $r=0$, $\cos\theta=-l/a$. And, since the space-time is asymptotically flat at conformal infinity where $\Omega=0$, the range of $r$ must be given by $r\in(0,r_\infty)$ where 
 \begin{equation}
 r_\infty= \left\{ 
 \begin{array}{cl}
 {\displaystyle {\omega\over\alpha(l+a\cos\theta)}} \qquad &\hbox{if} \quad a\cos\theta>-l \\[6pt]
 \infty \qquad &\hbox{otherwise}
 \end{array} \right. 
 \label{rangeragtl}
 \end{equation} 
 In the alternative case in which $a^2<l^2$ there is no curvature singularity and the range of $r$ can also take negative values. However, the coordinate $r$ does not extend to conformal infinity in all directions (i.e. for all values of $\theta$) when $\alpha\ne0$.  When $l>|a|$, \ $r\in(-\infty,r_\infty)$, \ where $r_\infty=\omega\alpha^{-1}(l+a\cos\theta)^{-1}$. And, when $l<-|a|$, $r\in(-|r_{\infty}|,\infty)$. In these cases, the space-time is not asymptotically flat at infinite values of $r$, as this does not correspond to conformal infinity. This is a natural feature of accelerating coordinates.

As fully described in \cite{GriPod05}, conical singularities generally occur on the axis. However, by specifying the range of $\phi$ appropriately, the singularity on one half of the axis can be removed. For example, that on $\theta=\pi$ is removed by taking $\phi\in\big[0,2\pi(1+a_3-a_4)^{-1}\big)$. In this case, the acceleration of the ``source'' is achieved by a ``string'' of deficit angle 
  \begin{equation}
 \delta_0 = {4\pi\,a_3 \over1+a_3-a_4} 
  \label{def0}
  \end{equation} 
 connecting it to infinity. Alternatively, the singularity on $\theta=0$ could be removed by taking $\phi\in\big[0,2\pi(1-a_3-a_4)^{-1}\big)$, and the acceleration would then be achieved by a ``strut'' between the sources in which the excess angle is given by 
  \begin{equation}
 -\delta_\pi = {4\pi\,a_3 \over1-a_3-a_4}. 
  \label{defpi}
  \end{equation} 
 The expressions (\ref{def0}) or (\ref{defpi}) are closely related to the tension or stress in the string or strut respectively and these should be equal for any given acceleration $\alpha$, at least according to Newtonian theory. However, the deficit/excess angles are the same fractions of the range of the periodic coordinate in each case. This, presumably, corresponds to an equality of forces in the general case.

It should also be recalled that, when $l\ne0$, the metric (\ref{newMetric}) has an additional singularity on $\theta=\pi$ which corresponds to the ``axis'' between two causally separated ``sources''. However, this can be switched to the other half-axis by the transformation $t'=t-4l\phi$. It can thus be seen that the topological singularity on the axis which causes the acceleration, and the singularity on the axis associated with the NUT parameter and the existence of closed timelike lines (see e.g. \cite{Bonnor69}), are mathematically independent. They may each be set on whatever parts of the axis may be considered to be most physically significant.

Intriguingly, there exists a possibility that the conical singularity may vanish on both halves of the axis simultaneously. This occurs when $a_3=0$: i.e. when
 $$ 2\alpha^2 l(\omega^2k+e^2+g^2) -{\textstyle{2\over3}}\omega^2l\Lambda
 =\alpha\omega m. $$ 
 However, this generally\footnote{
         It may be observed that this condition for a completely regular axis may also be satisfied when \ $a=0$, \ $l=\omega=0$ \ and \ $2\alpha(e^2+g^2)=m$. \ This is the special case of the extremely and oppositely charged $C$-metric. }
         only occurs when the NUT parameter $l$ is non-zero and, in this case, one of the ``half-axes'' is singular in a different way and is surrounded by a region containing closed timelike lines. Thus, the presence of charges or a cosmological constant is not sufficient to cause the sources to accelerate: some string-like structure is required (at least within this family of solutions).

Some special cases will now be considered in more detail. Namely, the non-accelerating solutions and those with no NUT parameter. The general case with $\Lambda=0$ has been described in detail in \cite{GriPod05} for which the charge-free case is summarised in figure~\ref{AccRotBHstructure}. Their particular subcases are well known. For the case with $a=0$, $\alpha$ is a redundant parameter, indicating that there are no accelerating NUT solutions without rotation.

\subsection{The Kerr--Newman--NUT--de~Sitter solution ($\alpha=0$)}
\label{KNNUTdS}

When $\alpha=0$, (\ref{k}) becomes \ $\omega^2k=(1-l^2\Lambda)(a^2-l^2)$ \ and hence (\ref{epsilon}) and (\ref{n}) become
 $$ \epsilon=1-({\textstyle{1\over3}}a^2+2l^2)\Lambda \,, \qquad\qquad n=l+{\textstyle{1\over3}}(a^2-4l^2)l\Lambda. $$ 
 The metric is then given by (\ref{newMetric}) with 
  $$ \begin{array}{l}
  \Omega=1 \\[6pt]
  \rho^2 =r^2+(l+a\cos\theta)^2 \\[6pt]
  \tilde P=\sin^2\theta(1+{\textstyle{4\over3}}\Lambda al\cos\theta +{\textstyle{1\over3}}\Lambda a^2\cos^2\theta) \\[6pt]
  Q= (a^2-l^2+e^2+g^2) -2mr + r^2 
  -\Lambda\Big[(a^2-l^2)l^2+({1\over3}a^2+2l^2)r^2+{1\over3}r^4\Big].
  \end{array} $$ 
 This is exactly the Kerr--Newman--NUT--de~Sitter solution in the form which is regular on the half-axis $\theta=0$. It represents a non-accelerating black hole with mass~$m$, electric and magnetic charges $e$ and $g$, a rotation parameter $a$ and a NUT parameter~$l$ in a de~Sitter or anti-de~Sitter background. It reduces to known forms when $l=0$ or $a=0$ or $\Lambda=0$.

In this Kerr--Newman--NUT--de~Sitter solution, it is important to distinguish the two cases in which $|a|$ is greater or less than $|l|$, as indicated in figure~\ref{AccRotBHstructure}. When $a^2\ge l^2$, $k\ge0$, the metric has a Kerr-like ring singularity at $r=0$, $\cos\theta=-l/a$, and the range of $r$ is given by (\ref{rangeragtl}). This case represents a Kerr--Newman--de~Sitter solution (a charged black hole) with a small NUT parameter. Alternatively, when $a^2<l^2$, $k<0$, the metric is singularity free, and the range of $r$ includes negative values. This case is best described as a charged NUT--de~Sitter solution with a small Kerr-like rotation. Although these two cases have identical metric forms, their singularity and global structures differ substantially.

\subsection{Accelerating Kerr--Newman--de~Sitter black holes ($l=0$)}

In the physically most significant case in which $\alpha$ is arbitrary but $l=0$, (\ref{k}) implies that \ $\omega^2k=a^2$. \ It is then convenient to use the remaining scaling freedom to put $\omega=a$, and hence 
 $$ \epsilon=1-\alpha^2(a^2+e^2+g^2)-{\textstyle{1\over3}}\Lambda a^2, \qquad k=1, \qquad n=-\alpha am. $$ 
 For this case, it can be seen explicitly that the Pleba\'nski--Demia\'nski parameter $n$ is non-zero, while the NUT parameter $l$ vanishes. The metric (\ref{newMetric}) now takes the form 
  \begin{equation}
 \d s^2={1\over\Omega^2}\left\{
{Q\over\rho^2}\left[\d t- a\sin^2\theta\,\d\phi \right]^2
   -{\rho^2\over Q}\,\d r^2  -{\tilde P\over\rho^2} \Big[ a\d t  
  -(r^2+a^2)\d\phi \Big]^2  
-{\rho^2\over\tilde P}\sin^2\theta\,\d\theta^2 \right\}, 
  \label{lzeroMetric}
  \end{equation}
  where
  $$ \begin{array}{l}
  \Omega=1-\alpha r\cos\theta \\[6pt]
  \rho^2 =r^2+a^2\cos^2\theta \\[6pt]
  \tilde P= \sin^2\theta \Big(1-2\alpha m\cos\theta +\big[\alpha^2(a^2+e^2+g^2)+{1\over3}\Lambda a^2\big]\cos^2\theta\Big) \\[6pt]
  Q= \Big((a^2+e^2+g^2) -2mr +r^2\Big) (1-\alpha^2r^2)-{1\over3}\Lambda(a^2+r^2)r^2.
  \end{array} $$ 
  The only non-zero components of the curvature tensor are given by
 $$  \begin{array}{l}
 {\displaystyle \Psi_2= \left(-m(1-i\alpha a)
+(e^2+g^2) {1+\alpha r\cos\theta\over r-ia\cos\theta} \right)
 \left({1-\alpha r\cos\theta\over r+ia\cos\theta}\right)^3 } , \\[12pt]
 {\displaystyle \Phi_{11}= {1\over2}\,(e^2+g^2)\,{(1-\alpha r\cos\theta)^4\over(r^2+a^2\cos^2\theta)^2}
 \qquad \hbox{and} \qquad \Lambda.}
  \end{array} $$ 
  These indicate the presence of a Kerr-like ring singularity at $r=0$, $\theta={\pi\over2}$. Thus, we may take the range of $r$ as 
 $$ r\in(0,r_\infty) \qquad \hbox{where} \qquad
 r_\infty= \left\{ 
 \begin{array}{cl}
 \alpha^{-1}\sec\theta \qquad &\hbox{if} \quad \theta<\pi/2 \\[4pt]
 \infty \qquad &\hbox{otherwise}
 \end{array} \right. $$

When $\Lambda=0$, the metric (\ref{lzeroMetric}) corresponds precisely to that of Hong and Teo \cite{HongTeo05} (and described in \cite{GriPod05}) which represents an accelerating and rotating black hole without any NUT-like behaviour and in which the acceleration is characterized by $\alpha$. In this case, if $m^2\ge a^2+e^2+g^2$, the expression for $Q$ factorises as 
  $$  Q = (r_--r)(r_+-r)(1-\alpha^2r^2), $$ 
  where
  \begin{equation}
  r_\pm = m\pm\sqrt{m^2-a^2-e^2-g^2}.
  \label{KerrNewman roots}
  \end{equation}
 The expressions for $r_\pm$ are identical to those for the locations of the outer (event) and inner (Cauchy) horizons of the non-accelerating Kerr--Newman black hole. However, in this case, there is another horizon at $r=\alpha^{-1}$ which is already familiar in the context of the $C$-metric as an acceleration horizon. For the case in which $\Lambda\ne0$, the locations of the horizons are modified.

It may also be observed that 
 $$ a_3=2\alpha m, \qquad a_4=-\alpha^2(a^2+e^2+g^2) -{\textstyle{1\over3}}\Lambda a^2, $$ 
 so that the deficit angle of the string causing the black holes to accelerate, or the excess angle of the strut between them, is obtained immediately using (\ref{def0}) or (\ref{defpi}) respectively. There are no closed timelike lines near the axis, confirming that this is the appropriate metric for describing a pair of accelerating and rotating black holes\footnote{
     It should be emphasised that this solution, which has no NUT-like properties, is different from the one that is obtained by putting $n=0$ and which is usually called the ``spinning $C$-metric''. This case, whose properties have been described in \cite{FarZim80b}--\cite{Pravdas02} etc. (at least when $e=g=0$ and $\Lambda=0$) still retains NUT-like properties such as the existence of closed timelike lines near one half of the axis. }
     as argued by Hong and Teo \cite{HongTeo05} for the case when $\Lambda=0$.

The metric (\ref{lzeroMetric}) nicely represents the singularity and horizon structure of an accelerating charged and rotating black hole in a de~Sitter or anti-de~Sitter background (see \cite{PodGri06}). It represents the space-time from the singularity through the inner and outer black hole horizons and out to and beyond the acceleration horizon. However, it does not cover the complete analytic extension inside the black hole horizon. For this, Kruskal--Szekeres-like coordinates are required as described for example for particular cases in \cite{HawEll73}. Neither does it cover the complete analytic extension beyond the acceleration horizon. This requires a transformation to boost-rotation-symmetric coordinates as has been given for the general case with $\Lambda=0$ in~\cite{GriPod06}. This shows that the complete space-time actually contains two causally separated charged and rotating black holes which accelerate away from each other in opposite spatial directions.

\subsubsection{The charged $C$-metric with a cosmological constant}
\label{section 4.3.1}

For the case in which $a=0$, the metric (\ref{lzeroMetric}) reduces to the simple diagonal form 
 $$ \d s^2={1\over(1-\alpha r\cos\theta)^2} \left( {Q\over r^2}\,\d t^2 -{r^2\over Q}\,\d r^2
  -\tilde P\,r^2\,\d\phi^2
 -{r^2\sin^2\theta\over\tilde P}\,\d\theta^2 \right), $$ 
 where 
  $$ \begin{array}{l}
  \tilde P=\sin^2\theta\Big(1-2\alpha m\cos\theta +\alpha^2(e^2+g^2)\cos^2\theta\Big), \\[6pt]
  Q=(e^2+g^2-2mr+r^2)(1-\alpha^2r^2)-{1\over3}\Lambda r^4.
  \end{array} $$ 
 When $\Lambda=0$, this is exactly equivalent to the form for the charged $C$-metric that was introduced recently by Hong and Teo \cite{HongTeo03} using the coordinates \ $x=-\cos\theta$ \ and \ \hbox{$y=-1/(\alpha r)$}. \ It describes a pair of black holes of mass $m$ and electric and magnetic charges $e$ and $g$ which accelerate towards infinity under the action of forces represented by a conical singularity, for which $\alpha$ is precisely the acceleration. In this case, the acceleration horizon at $r=\alpha^{-1}$ is clearly identified. However, when $\Lambda\ne0$, the location of all horizons is modified.

Further properties of the charged $C$-metric in a (anti-)de~Sitter background have been analysed in \cite{PodGri01} and \cite{DiaLem03a}--\cite{PoOrKr03}. However, these worked with a different form of the metric to that above.

\section{Twisting but non-accelerating solutions ($\alpha=0$)}
\label{nonaccSolns}

Let us now return to the general family of expanding type~D solutions given by the metric (\ref{altPlebMetric}) but concentrate here on the particular case for which \ $\alpha=0$, \ implying that \ $\Omega=1$. \ The line element is thus 
  \begin{equation}
  \begin{array}{l}
{\displaystyle \d s^2=
{Q\over\rho^2}\left[\d t- \left(a(1-\tilde p^2)
+2l(1-\tilde p) \right)\d\phi \right]^2
   -{\rho^2\over Q}\,\d r^2 
} \\[8pt]
  \hskip8pc {\displaystyle 
 -{\tilde P\over\rho^2} \Big[ a\d t  
  -\Big(r^2+(l+a)^2\Big)\d\phi \Big]^2  
-{\rho^2\over\tilde P}\,\d\tilde p^2 , }
\end{array}
  \label{nonaccMetric}
  \end{equation}
  where
  $$ \begin{array}{l}
  \rho^2 =r^2+(l+a\tilde p)^2 \\[6pt]
  \tilde P= a_0 +a_1\tilde p +a_2\tilde p^2
+ a_3\tilde p^3 +a_4\tilde p^4 \\[8pt]
  Q= (\omega^2k+e^2+g^2) -2mr +\epsilon r^2 
-{1\over3}\Lambda r^4,
  \end{array} $$   
  and 
  \begin{equation}
 \begin{array}{l}
  a_0= a^{-2} \Big( \omega^2k +2nl -\epsilon l^2
-{1\over3}\Lambda l^4 \Big) \\[6pt]
  a_1=2a^{-1}\Big( n -\epsilon l
-{2\over3}\Lambda l^3 \Big) \\[6pt]
  a_2= -\epsilon -2\Lambda l^2 \\[6pt]
  a_3= -{4\over3}\Lambda al \\[6pt]
  a_4= -{1\over3}\Lambda a^2.
  \end{array} 
  \label{ainonacc}
  \end{equation} 
 As explained in section~\ref{expanding}, the properties of the solutions in this family depend critically on the character of the quartic $\tilde P(\tilde p)$. However, it may be noticed here that the physical parameters $m$, $e$ and $g$ do not appear explicitly in this function, but the remaining parameters $n$, $k$, $\epsilon$, $\Lambda$, $a$, $l$ and $\omega$ must be such that there exists at least one range of $\tilde p$ in which $\tilde P>0$. When more than one such range exists, the different possible ranges correspond to distinct space-times which have different physical properties.

Although it is generally possible to choose $l$ to put $a_1=0$, it is more convenient to choose $l$ to simplify the factors of $\tilde P$. In fact, when $\Lambda a=0$, $\tilde P$ is a quadratic which can be put into one of its distinct canonical forms. Otherwise, $\tilde P$ can be expressed as the product of two quadratics, one of which can be similarly simplified.

\subsection{Non-accelerating cases with $\Lambda=0$} 

When both $\alpha=0$ and $\Lambda=0$, $\tilde P$ reduces to a quadratic and we obtain the line element (\ref{nonaccMetric}) in which 
  $$ \begin{array}{l}
  \rho^2 =r^2+(l+a\tilde p)^2 \\[6pt]
  \tilde P= a_0 +a_1\tilde p -\epsilon\tilde p^2 \\[8pt]
  Q= (\omega^2k+e^2+g^2) -2mr +\epsilon r^2 ,
  \end{array} $$   
  with 
  $$ a_0= a^{-2} \Big( \omega^2k +2nl -\epsilon l^2 \Big), \qquad
  a_1=2a^{-1}\Big( n -\epsilon l \Big) . $$
 In this metric, $a$ and $l$ can be chosen for convenience and, in addition, we still have two scaling freedoms. The resulting family of space-times generally has five arbitrary parameters $m$, $n$ (preferably expressed in terms of $l$), $e$, $g$ and $a$ (replacing $\omega$).

It is now possible to fix $a$ and $l$ (equivalent to the linear transformation (\ref{trans1A}) in $p)$ to put $\tilde P$ (which is necessarily positive) into one of its six distinct canonical forms, namely 
 $$ 1-\tilde p^2, \qquad \tilde p^2+1, \qquad \tilde p^2-1, \qquad \tilde p^2, \qquad \tilde p, \qquad 1. \leqno{\qquad\tilde P\,:} $$ 
 When the electromagnetic field vanishes, this yields the family of class~II solutions identified by Kinnersley \cite{Kinnersley69}. Let us now consider each of these possibilities explicitly.

When $\epsilon>0$, we can put \ $\epsilon=1$ \ and set \ $l=n$ \ so that \ $a_1=0$. \ We can then set \ $\omega^2k=a^2-l^2$, \ so that \ $a_0=1$, \ and thus \ $\tilde P=1-\tilde p^2$. This case corresponds exactly the Kerr--Newman--NUT solution described in subsection~\ref{KNNUTdS} above, in which it is natural to put \ $\tilde p=\cos\theta$. \ In the vacuum case, it is the Kerr--NUT solution, which is the Kinnersley class~II.A solution.

When $\epsilon<0$, we can put \ $\epsilon=-1$ \ and set \ $l=-n$ \ so that again \ $a_1=0$. \ We can then set \ $\omega^2k=l^2+\epsilon_0a^2$, \ where \ $\epsilon_0=1,-1,0$, \ so that \ $\tilde P=\tilde p^2+\epsilon_0$ \ and \ $Q=(e^2+g^2+l^2+\epsilon_0a^2)-2mr-r^2$. \ For the vacuum case and with different values of $\epsilon_0$, these correspond to the Kinnersley class II.B, II.C and II.D solutions respectively. (When \ $\epsilon_0=1$ \ it is convenient to put \ $\tilde p=\sinh\theta$, \ and when \ $\epsilon_0=-1$ \ it is convenient to put \ $\tilde p=\cosh\theta$.)

Finally, we consider the case in which $\epsilon=0$. When $n\ne0$, it is convenient to set \ $a=2n$ \ and \ $\omega^2k=-2nl$ \ so that \ $a_1=1$ \ and \ $a_0=0$. \ With these choices, the metric functions become \ $\tilde P=\tilde p$ \ and \ $Q=(e^2+g^2-al)-2mr$. \ For the vacuum case, this is the Kinnersley class II.E solution. Alternatively, for the case in which $n=0$, it is convenient to set \ $\omega^2k=a^2$ \ so that \ $a_0=1$ \ and the metric functions become \ $\tilde P=1$ \ and \ $Q=(e^2+g^2+a^2)-2mr$. \ For the vacuum case, this is of Kinnersley class II.F.

\subsection{Non-accelerating cases with $\Lambda\ne0$ and $n=0$} 

When both $\alpha=0$ and $n=0$, it is appropriate to set $l=0$ so that $\tilde P$ is a quadratic in $\tilde p^2$. In this case, we obtain the metric (\ref{nonaccMetric}) with 
  $$ \begin{array}{l}
  \rho^2 =r^2+a^2\tilde p^2 \\[6pt]
  \tilde P= a^{-2} \omega^2k -\epsilon\tilde p^2 
  -{1\over3}\Lambda a^2\tilde p^4 \\[8pt]
  Q= (e^2+g^2+\omega^2k) -2mr +\epsilon r^2 
-{1\over3}\Lambda r^4.
  \end{array} $$ 
 Various particular cases now need to be considered. For different values of the essential parameters, the free parameters have to be chosen to obtain a convenient range of $\tilde p$ in which $\tilde P>0$. Here, we will only consider $\Lambda\ne0$, since all the cases with $\Lambda=0$ are already described in the previous subsection.

\subsubsection{The Kerr--Newman--de~Sitter solution}

When \ $k>0$ \ we may choose $\omega^2k=a^2$ \ and then set \ $\epsilon=1-{1\over3}\Lambda a^2$ \ provided the signs are consistent. Thus we obtain  
 $$ \begin{array}{l}
 \tilde P =(1-\tilde p^2)(1+{1\over3}\Lambda a^2\,\tilde p^2) \\[8pt]
 Q=(e^2+g^2+a^2) -2mr +(1-{1\over3}\Lambda a^2)r^2 
-{\textstyle{1\over3}}\Lambda r^4 \\[8pt]
 \quad =(e^2+g^2) -2mr +(r^2+a^2)(1-{1\over3}\Lambda r^2) 
 \end{array} $$ 
 which is obviously the $l=0$ subcase of the space-times discussed in subsection~\ref{KNNUTdS}. The metric can then be written in the standard form of the Kerr--Newman--de~Sitter solution in Boyer--Lindquist-type coordinates \cite{Carter70}, \cite{GibHaw77} using the transformation 
 \begin{equation}
 \tilde p=\cos\theta, \qquad t=\bar t\,\Xi^{-1}, 
 \qquad \phi=\bar\phi\,\Xi^{-1},
 \label{KerrdSTrans}
 \end{equation}  
 where $\Xi=1+{1\over3}\Lambda a^2$. This puts the metric in the form 
 \begin{equation}
 \d s^2= {\Delta_r\over\Xi^2\rho^2}
 \Big[\d\bar t-a\sin^2\theta\,\d\bar\phi\Big]^2 
 -{\Delta_\theta\sin^2\theta\over\Xi^2\rho^2}\Big[a\d\bar t-(r^2+a^2)\d\bar\phi
\Big]^2 
 -{\rho^2\over \Delta_r}\,\d r^2 
 -{\rho^2\over\Delta_\theta}\,\d\theta^2 ,
 \label{KerrNdS}
 \end{equation} 
 where 
 \begin{equation}
 \begin{array}{l}
 \rho^2=r^2+a^2\cos^2\theta \,, \\[6pt]
 \Delta_r =(r^2+a^2)(1-{1\over3}\Lambda r^2) -2mr +(e^2+g^2) , \\[6pt]
 \Delta_\theta=1+{1\over3}\Lambda a^2\cos^2\theta \,.
 \end{array}
 \end{equation} 
 Formally, there is no need to introduce the constant scaling $\Xi$ in $t$ and $\phi$. However, this is included so that the metric has a well-behaved axis at $\theta=0$ and $\theta=\pi$ with $\phi\in[0,2\pi)$.

It should also be noted that, if $\Lambda<0$, \ $\tilde P(\tilde p)=0$ \ has four roots: \ $\tilde p=\pm1$ \ and \ 
$\tilde p=\pm\sqrt{3\over-\Lambda}\,{1\over a}$. \ If \ ${1\over3}\Lambda a^2>-1$, \ the above solution is valid for \ $0\le\theta\le\pi$. \ However, if \ ${1\over3}\Lambda a^2<-1$, \ the solution will only be valid for the smaller range of $\theta$ for which \ $\Delta_\theta\ge0$ \ and the space-time in this case does not represent a black hole in an anti-de~Sitter background.

\subsubsection{Other cases}   

When $\epsilon=0$ and $\Lambda>0$, it is necessary that $k>0$. In this case,
$\tilde p$ varies between the two real roots of~$\tilde P$. However, when
$\Lambda<0$, $k$ can take any value and $\tilde p$ will take various appropriate
ranges which all extend to~$\pm\infty$.

When $\epsilon\ne0$ and $k>0$, we may choose $\omega^2k=a^2$. The
Kerr--Newman--de~Sitter solution is obtained as above when the sign of
$\epsilon$ is such that it is possible to make a rescaling to set \
$\epsilon=1-{1\over3}\Lambda a^2$. \ However, if the signs are inconsistent, it
is always possible to use the scaling to set \ $\epsilon=-1+{1\over3}\Lambda
a^2$. \ In this case, we obtain \ $\tilde P=(1+\tilde p^2)(1-{1\over3}\Lambda
a^2\,\tilde p^2)$. \ If $\Lambda<0$, this is a valid solution for all $\tilde
p\in(-\infty,\infty)$. However, if $\Lambda>0$, it is valid for $\tilde
p\in[-\sqrt{3\over\Lambda}\,{1\over a},\sqrt{3\over\Lambda}\,{1\over a}]$.

When $\epsilon\ne0$ and $k<0$, we can choose $\omega^2k=-a^2$ and set $\epsilon=-1-{1\over3}\Lambda a^2$ provided the signs are consistent. In this case \ $\tilde P=(\tilde p^2-1)(1-{1\over3}\Lambda\tilde a^2\tilde p^2)$. \ We can then choose $\tilde p=\cosh\theta$, so that \
$\Delta_r =-(r^2+a^2)(1+{1\over3}\Lambda r^2) -2mr +(e^2+g^2)$ \ and \ 
$\Delta_\theta=1-{1\over3}\Lambda a^2\cosh^2\theta$. \ However, even this is only valid over the entire range of $\theta$ if $\Lambda<0$. \
If \ $0<\Lambda<3a^{-2}$, \ $\tilde P$ generally has four roots and two distinct ranges of $\tilde p$ are possible. And if $\Lambda>3a^{-2}$, we can again put $\tilde p=\cos\theta$.

If the signs are inconsistent in the above case for which $\epsilon\ne0$ and
$k<0$, we can set $\epsilon=1+{1\over3}\Lambda a^2$. In this case, \ 
$\tilde P=-(1+\tilde p^2)(1+{1\over3}\Lambda\tilde a^2\tilde p^2)$. \ Of course, this is only positive if $\Lambda<0$ and $\tilde p^2$ is sufficiently large.

Finally, we must consider the remaining case in which $\epsilon\ne0$ and $k=0$. In this case, \ $\tilde P=-\tilde p^2(\epsilon +{1\over3}\Lambda a^2\tilde p^2)$ \ which is only permitted if $\epsilon$ and $\Lambda$ are not both positive. If at least one of these parameters is negative appropriate ranges of $\tilde p$ can easily be obtained.

\subsection{Non-accelerating cases with $\Lambda\ne0$ and $a=0$} 
\label{nonaczeroa}

Let us now consider the complementary case in which both $\alpha=0$ and $a=0$. However, the coefficients of $\tilde P$ given in (\ref{ainonacc}) must be bounded in the limit as $a\to0$. We thus set  
  $$ \begin{array}{rll}
 \omega^2k +2nl -\epsilon l^2 -{1\over3}\Lambda l^4 &=& a_0a^2 \\[6pt]
 n -\epsilon l -{2\over3}\Lambda l^3 &=& {1\over2}a_1a \\[6pt]
 \epsilon +2\Lambda l^2 &=& -a_2
  \end{array} $$ 
 and then take the limit $a\to0$, so that $a_3=0=a_4$. Thus, the metric function $\tilde P$ in (\ref{nonaccMetric}) is again the quadratic 
  $$ \tilde P= a_0 +a_1\tilde p +a_2\tilde p^2. $$ 
 We also obtain 
 $$ \begin{array}{rll}
\epsilon &=& -a_2-2\Lambda l^2, \\[6pt]
n &=& -a_2l-{4\over3}\Lambda l^3 , \\[6pt]
\omega^2k &=& a_2 l^2 +\Lambda l^4 , 
 \end{array} $$  
 and the metric (\ref{nonaccMetric}) can be written as 
 \begin{equation} 
 \d s^2= \tilde Q\Big(\d t
-2l(1-\tilde p)\,\d\phi\Big)^2 -{\d r^2\over\tilde Q} 
 -(r^2+l^2)\Big(\tilde P\,\d\phi^2 
+{1\over\tilde P}\,\d\tilde p^2 \Big)
 \label{lambda0} 
 \end{equation}  
 where 
 $$ \tilde Q= {1\over r^2+l^2} \left[ a_2(l^2-r^2) +(e^2+g^2) -2mr
 -\Lambda\left({\textstyle{1\over3}}r^4 +2l^2 r^2 - l^4 \right)
 \right]. $$ 
 In this form, and with appropriate values of $a_0$ that will be clarified below, it may immediately be recognised that the three solutions with
different values of $a_2$ correspond exactly to the general family of NUT
solutions with additional non-zero electric and magnetic charges and a non-zero
cosmological constant.

We now consider all possible distinct subcases according to the usual canonical forms of $\tilde P$ in which the coefficients $a_i$ take the values $1$, $-1$ or $0$.

\subsubsection{The Taub--NUT--de~Sitter solution}

In the case in which $a_0=1$, $a_1=0$ and $a_2=-1$ so that 
$\tilde P=1-\tilde p^2$ and $|\tilde p|\le1$, we can put $\tilde p=\cos\theta$ so that the metric (\ref{lambda0}) can be written as 
 \begin{equation} 
 \d s^2= \tilde Q\Big(\d t
-4l\sin^2{\textstyle{\theta\over2}}\,\d\phi\Big)^2 -{\d r^2\over\tilde Q} 
 -(r^2+l^2)\Big(\d\theta^2+\sin^2\theta\,\d\phi^2 \Big)
 \label{TaubNUT} 
 \end{equation}  
 where  
 $$ \tilde Q= {1\over r^2+l^2} \left[ r^2-2mr-l^2 +e^2+g^2 
 -\Lambda\left({\textstyle{1\over3}}r^4 +2l^2 r^2 -l^4 \right)
 \right]. $$ 
 This is exactly the known extension of the Taub--NUT metric to include non-zero charges and a cosmological constant \cite{Ruban72}, in which $l$ is the NUT parameter. It is contained in the case described in subsection~\ref{KNNUTdS} for~$a=0$.

\subsubsection{Other cases}
\label{TNUTcases}

There are five remaining cases to be considered. However, it is generally convenient here to set $t=t'+2l\phi$.  

\medskip
For the case when $a_0=1$, $a_1=0$ and $a_2=0$, in which $\tilde P=1$, the metric (\ref{lambda0}) becomes 
 \begin{equation} 
 \d s^2= \tilde Q\Big(\d t'
+2l\,\tilde p\,\d\phi\Big)^2 -{\d r^2\over\tilde Q} 
 -(r^2+l^2)\Big(\d\phi^2 +\d\tilde p^2 \Big),
 \label{e2zero} 
 \end{equation}  
 where 
 \begin{equation} 
 \tilde Q= {1\over r^2+l^2} \left[ -2mr +e^2+g^2
 -\Lambda\left({\textstyle{1\over3}}r^4 +2l^2 r^2 -l^4 \right)
 \right]. 
 \label{Qa2zero} 
 \end{equation}

\medskip
For the case when $a_0=-1$, $a_1=0$ and $a_2=1$, in which
$\tilde P=\tilde p^2-1$ and $|\tilde p|\ge1$, we put \ $\tilde p=\cosh R$ \ so that the metric (\ref{lambda0}) becomes 
 \begin{equation} 
 \d s^2= \tilde Q\Big(\d t'
+2l\cosh R\,\d\phi\Big)^2 -{\d r^2\over\tilde Q} 
 -(r^2+l^2)\Big(\d R^2+\sinh^2R\,\d\phi^2 \Big)
 \label{NUT3em1} 
 \end{equation}  
 where 
 \begin{equation} 
 \tilde Q= {1\over r^2+l^2} \left[ -r^2-2mr+l^2 +e^2+g^2 
 -\Lambda\left({\textstyle{1\over3}}r^4 +2l^2 r^2 -l^4 \right) \right]. 
 \label{Qa2one} 
 \end{equation} 
 However, in this case, it would have been more convenient to revert to the coordinate $t$. With this, the metric would be regular at $R=0$. This would then corresponds to an axis, so that $\phi$ may be interpreted as an angular coordinate.

\medskip 
The above two cases, together with the case $a_0=1$, $a_1=0$ $a_2=-1$ that was described in the previous subsection, constitute the three members of the general family of NUT (Kinnersley's class~I) solutions when $e=g=\Lambda=0$. They thus generalise these solutions to include non-zero electric and magnetic charges and a non-zero cosmological constant. They correspond, in the order given above, to non-rotating ($a=0$) subcases of the solutions of Kinnersley's classes II.A, II.F and II.C generalized to include charges and a cosmological constant. However, for the family of solutions being discussed here, there are apparently still three additional cases to be considered.

\medskip\goodbreak
Let us now consider the case in which $a_0=0$, $a_1=1$ and $a_2=0$. In this case, $\tilde P=\tilde p$, and the metric takes the form 
 $$ \d s^2= \tilde Q\Big(\d t' 
+2l\,\tilde p\,\d\phi\Big)^2 -{\d r^2\over\tilde Q} 
 -(r^2+l^2)\Big(\tilde p\,\d\phi^2 
+{1\over\tilde p}\,\d\tilde p^2 \Big) $$ 
 where $Q$ is given by (\ref{Qa2zero}). Putting $\tilde p={1\over4}\rho^2$ and
$\phi=2\psi$, the metric becomes 
 \begin{equation} 
 \d s^2= \tilde Q\Big(\d t' +l\,\rho^2\d\phi\Big)^2 -{\d r^2\over\tilde Q} 
 -(r^2+l^2)\Big(\d\rho^2 +\rho^2\,\d\psi^2 \Big) \,.
 \label{e2e0zero} 
 \end{equation}  
 This is a non-rotating ($a=0$) solution of Kinnersley's class II.E with additional parameters. However, the further transformation \ $t'=\tau+lxy$, \ $\rho=\sqrt{x^2+y^2}$, \ $\psi=\tan^{-1}(y/x)$ \ transforms (\ref{e2e0zero}) to a form which is identical to that of (\ref{e2zero}) above. Thus, in
this non-rotating limit, solutions of Kinnersley's classes II.E and II.F are
equivalent even with additional charges and a cosmological constant.

\medskip\goodbreak
For the case when $a_2=1$, $a_1=0$ and $a_0=0$, in which
$\tilde P=\tilde p^2$, we put $\tilde p=e^\chi$ so that the metric
(\ref{lambda0}) becomes 
 \begin{equation} 
 \d s^2= \tilde Q\Big(\d t'
+2l\,e^\chi\d\phi\Big)^2 -{\d r^2\over\tilde Q} 
 -(r^2+l^2)\Big(\d\chi^2+e^{2\chi}\d\phi^2 \Big)
 \label{NUT3e0} 
 \end{equation}  
 where $Q$ is given by (\ref{Qa2one}). This is a non-rotating ($a=0$) solution of Kinnersley's class II.D with additional parameters. However, the transformation 
 $$ e^\chi=\cosh R+\sinh R\cos\psi, \qquad 
\phi={\sin\psi\over\cos\psi+\coth R} $$ 
 together with 
 $$ t'=\tau +2l\psi 
+4l\tan^{-1}\left( {\tanh({R\over2})+\cos\psi\over\sin\psi} \right) $$ 
 takes the metric (\ref{NUT3e0}) exactly to the form (\ref{NUT3em1}).

\medskip\goodbreak
Finally, for the case when $a_2=1$, $a_1=0$ and $a_0=1$, in which $\tilde P=1+\tilde p^2$, we put $\tilde p=\sinh\psi$ so that the metric
(\ref{lambda0}) becomes 
 \begin{equation} 
 \d s^2= \tilde Q\Big(\d t'
+2l\sinh\psi\,\d\phi\Big)^2 -{\d r^2\over\tilde Q} 
 -(r^2+l^2)\Big(\d\psi^2+\cosh^2\psi\,\d\phi^2 \Big)
 \label{NUT3e1} 
 \end{equation}  
 where $Q$ is given by (\ref{Qa2one}). This is a non-rotating ($a=0$) solution of Kinnersley's class II.B with additional parameters. However, applying the transformation 
 $$ \begin{array}{l}
 {\displaystyle e^\psi=\sinh\chi+\cosh\chi\cosh\sigma_B, \qquad 
\phi={\sinh\sigma\over\cosh\sigma+\tanh\chi}, } \\[12pt]
 {\displaystyle t'=\tau -2l\tan^{-1}(\sinh\sigma) 
 +4l\tan^{-1}\left( {1+\tanh({\chi\over2})\cosh\sigma\over\sinh\sigma}
\right) }
 \end{array} $$ 
 to the metric (\ref{NUT3e0}) takes it exactly to the above form (\ref{NUT3e1}). This confirms that these generalised solutions of Kinnersley's classes II.B, II.C and II.D (i.e. with $a_2=1$ and any value of $a_0$) become completely equivalent in the class~I solutions for which $a=0$.

It is therefore concluded that the family of charged NUT--de~Sitter solutions
which have five continuous parameters $m$, $l$, $e$, $g$ and $\Lambda$ come in
just three different varieties according to the values of the discrete parameter
$a_2=\pm1,0$. Of these, the case in which $a_2=-1$ may be referred to as the charged Taub--NUT--de~Sitter solution.

Finally, let us remark that when $l=0$, $e=g=0$ and $\Lambda=0$ ($m\ne0$), the three possible solutions are the so-called $A$-metrics \cite{EhlersKundt62}. When $a_2=-1$, it is the $AI$ metric which is the Schwarzschild solution. When $a_2=1$, it is the $AII$ metric which is its hyperbolic equivalent. And when $a_2=0$, it is the $AIII$ metric which is either the type~D Kasner solution or Taub's plane symmetric solution according to the sign of~$m$.

\section{Accelerating but non-twisting solutions ($\omega=0$)}

In this section we will consider the twist-free Pleba\'nski--Demia\'nski solutions (\ref{PleDemMetric}), (\ref{PQeqns}) in which $\omega=0$. In this case it is necessary that $n=0$, but $n'=\omega^{-1}n$ may be non-zero and the metric takes the diagonal form 
  \begin{equation}
  \d s^2={1\over(1-\alpha pr)^2} \Bigg[
 \tilde Q\,\d\tau^2 -{\d r^2\over\tilde Q}
  -r^2\left( {\d p^2\over P} +P\,\d\sigma^2 \right) \Bigg],
  \label{twistfreeMetric}
  \end{equation}
  where
 \begin{equation}
 \begin{array}{l}
 P =k +2n'p -\epsilon p^2 +2\alpha mp^3
-\alpha^2(e^2+g^2)p^4, \\[8pt]
 {\displaystyle \tilde Q = \epsilon -{2m\over r} +{e^2+g^2\over r^2} 
 -2\alpha n'r -(\alpha^2k+\Lambda/3)r^2 },
 \end{array}
 \label{twistfreePQeqns}
 \end{equation} 
 and $m$, $n'$, $e$, $g$, $\Lambda$, $\epsilon$, $k$ and $\alpha$ are arbitrary real parameters. (It is particularly significant here that the cosmological constant does not appear explicitly in the expression for~$P$.) These solutions are in fact the type~D Robinson--Trautman solutions. This follows from the observation that the repeated principal null congruences are geodesic, shear-free, twist-free and expanding.

When $\alpha=0$, $P$ is quadratic and canonical coordinates can be used. For example, when $\epsilon>0$, we can take $\epsilon=1$, remove $n'$ and set $k=1$. It is then natural to put \ $p=\cos\theta$ \ so that \ $P=\sin^2\theta$ \ and the resulting metric is the familiar form of the Reissner--Nordstr\"om--de~Sitter solution. For alternative canonical choices the equivalent zero and negative curvature solutions are obtained. These cases in which $\alpha=0$ are the $l=0$ subcases of the family of solutions described in section~\ref{nonaczeroa}.

When $\alpha\ne0$, we note that $P$ is generally a quartic. As such, it can always be written as the product of two quadratic factors. However, we can now consider the transformation which leaves the conformal factor in (\ref{twistfreeMetric}) invariant, namely
 \begin{equation}
 p=\beta+\tilde p, \qquad\qquad r= {\tilde r\over1+\alpha\beta\tilde r}\,, 
 \label{accfreedom}
 \end{equation} 
 where $\beta$ is a constant, Under this freedom, the metric (\ref{twistfreeMetric}) is unchanged with $e$, $g$ and $\Lambda$ also unchanged, but the other parameters are transformed as 
 \begin{equation}
 \begin{array}{l} 
 m= \tilde m-2\alpha\beta(e^2+g^2) \,, \\[6pt]
 \epsilon= \tilde\epsilon -6\alpha\beta\tilde m +6\alpha^2\beta^2(e^2+g^2)
 \,, \\[6pt]
 n'= \tilde n' -\beta\tilde\epsilon +3\alpha\beta^2\tilde m
-2\alpha^2\beta^3(e^2+g^2) \,, \\[6pt]
 k= \tilde k +2\beta\tilde n' -\beta^2\tilde\epsilon +2\alpha\beta^3\tilde m
-\alpha^2\beta^4(e^2+g^2) \,.
 \end{array}
 \label{parametertrans}
 \end{equation}
 Unless $e=g=0$ and $\epsilon^2<12\alpha ml'$, it is possible to use the freedom (\ref{accfreedom}) to remove the parameter $n'$. However, it is more useful in general to use this freedom, together with a rescaling of coordinates, to simplify one of the quadratic factors of~$P(p)$. In particular, we can introduce a new parameter $\epsilon'$ and use the available freedoms to set 
 $$ \epsilon=\epsilon'-{k\alpha^2(e^2+g^2)\over\epsilon'}, 
 \qquad n'=-{k\alpha m\over\epsilon'},  $$ 
 so that 
 $$ \begin{array}{l}
 {\displaystyle P=(k-\epsilon'p^2) \left(1-{2\alpha m\over\epsilon'}p +{\alpha^2(e^2+g^2)\over\epsilon'}p^2\right) }\,, \\[12pt]
 {\displaystyle \tilde Q = \left(1-{k\alpha^2\over\epsilon'}\,r^2\right)
 \left(\epsilon'-{2m\over r} +{e^2+g^2\over r^2}\right) 
 -{\Lambda\over3}\,r^2} \,, 
 \end{array} $$ 
 where $\epsilon'$ and $k$ can be scaled to appropriate values.

Consider first the case in which $\epsilon'$ and $k$ are both positive and scaled to unity: i.e. when
 $$ \epsilon=1-\alpha^2(e^2+g^2), \qquad k=1, \qquad n'=-\alpha m. $$ 
 In this case 
 $$ \begin{array}{l}
 P=(1-p^2)\Big(1-2\alpha mp+\alpha^2(e^2+g^2)p^2\Big)\,, \\[12pt]
 {\displaystyle \tilde Q = (1-\alpha^2r^2)
 \left(1-{2m\over r} +{e^2+g^2\over r^2}\right) 
 -{\Lambda\over3}\,r^2} \,,  
 \end{array} $$  
 which is precisely the form of the charged $C$-metric given in \cite{GriPod05} except that an arbitrary cosmological constant is now included (see also section~\ref{section 4.3.1}). This case thus represents a charged black hole which is accelerating in a de~Sitter, anti-de~Sitter or Minkowski background with acceleration $\alpha$, under the action of a string-like structure represented by a conical singularity.

Other space-times are obtained when $k$ and/or $\epsilon'$ are negative, or $k=0$, although these may have no physically significant interpretation. For these cases, the parameter $\alpha$ may not have the interpretation as an acceleration. As above, the range of $p$ must always be such that $P>0$. 
Thus, if $k/\epsilon'>0$, $p$ may have a limiting value of $\sqrt{k/\epsilon'}$ and/or $-\sqrt{k/\epsilon'}$. And if $m^2>\epsilon'(e^2+g^2)$, other finite limiting values of $p$ will also occur. If $\epsilon'<0$, the range of $p$ may extend to $\pm\infty$.

One obvious advantage of the line element (\ref{twistfreeMetric}) for the $C$-metric and its generalizations over more familiar forms is that it possesses well-behaved limits when $\alpha$ vanishes.

\newpage
\section{Non-expanding solutions} 
\label{nonexpsolns}

The above sections have explored all the various cases of the Pleba\'nski--Demia\'nski family in which the repeated principal null congruences have non-zero expansion. This has been achieved after first extending the metric (\ref{PleDemMetric}) to the form (\ref{altPlebMetric}) by including a shift in $p$. Our purpose now is to examine the alternative situation in which these congruences are non-expanding so that they are not covered by the metric (\ref{PleDemMetric}). And, since solutions with a non-expanding shear-free null geodesic congruence must be twist-free, the twist must also vanish in this case. The solutions obtained are therefore necessarily the type~D solutions of Kundt's class. (An associated family of conformally flat solutions will be considered later in section~\ref{confflat}).

Since these solutions are non-twisting, $\omega$ cannot be a twist parameter and, in general, we may put $\omega=1$. However, a special case occurs when $\omega=0$. This will be treated separately in section~\ref{nonexpomegazero} below.

\subsection{Another modified form of the metric}

For the case in which $\omega\ne0$, we start with the metric (\ref{PleDemMetric}), put $\omega=1$, and apply the transformation 
 \begin{equation} 
 r= \gamma+\kappa q, \qquad 
\sigma=\kappa^{-1}\,t,
\qquad \tau= \psi-\gamma^2\kappa^{-1}\,t. 
 \label{trans1B} 
 \end{equation}   
 where $\gamma$ and $\kappa$ are arbitrary parameters. (There is no need to include a shift in $p$ at this stage. The possibility of including such a transformation will be considered later.) The resulting line element is then 
  \begin{equation}
  \d s^2={1\over\Omega^2} \Bigg[
{\tilde Q\over\rho^2}\Big(\kappa\d\psi-(\gamma^2+p^2)\d t\Big)^2
  -{P\over\rho^2}
  \Big(\d\psi+(2\gamma q+\kappa q^2)\d t \Big)^2 
  -{\rho^2\over P}\,\d p^2 -{\rho^2\over\tilde Q}\,\d q^2 \Bigg], 
  \label{qtransMetric}
  \end{equation}
  where 
 $$ \begin{array}{l}
 \Omega=1-\alpha(\gamma+\kappa q)p \,, \\[8pt]
 \rho^2= (\gamma+\kappa q)^2+p^2 \,, \\[8pt]
 P= k +2np -\epsilon p^2 +2\alpha mp^3
-\big[\alpha^2(k+e^2+g^2)+\Lambda/3\big]p^4 \,, \\[8pt]
 \tilde Q= \epsilon_0+\epsilon_1q-\epsilon_2q^2 
 -(2\alpha n+4\alpha^2k\gamma +{4\over3}\Lambda\gamma)\kappa q^3
 -(\alpha^2k+{1\over3}\Lambda)\kappa^2q^4 \,,
 \end{array} $$ 
 and 
 $$ \begin{array}{l}
 \epsilon_0= \kappa^{-2}(k+e^2+g^2 -2m\gamma +\epsilon\gamma^2 
 -2\alpha n\gamma^3 -\alpha^2k\gamma^4 
-{1\over3}\Lambda\gamma^4) \,, \\[6pt] 
 \epsilon_1= 2\kappa^{-1}(-m +\epsilon\gamma -3\alpha n\gamma^2 -2\alpha^2k\gamma^3 -{2\over3}\Lambda\gamma^3) \,, \\[6pt] 
 \epsilon_2= -\epsilon +6\alpha n\gamma +6\alpha^2k\gamma^2 
 +2\Lambda \gamma^2 \,. 
 \end{array} $$ 
 We may make the simple observation at this point that the metric (\ref{qtransMetric}) explicitly includes both expanding type~D solutions when $\kappa\ne0$ and non-expanding solutions when $\kappa=0$.

\subsection{Non-expanding cases with $\omega\ne0$}

To obtain the general family of non-expanding solutions, we now put $\kappa=0$ in the above metric. (It can be seen from (\ref{spincoeffts}) that, since $Q(\gamma)\approx\kappa^2\epsilon_0$ here, the expansion and twist of the repeated null congruence must vanish in this limit.) 
For space-times to exist in this limit, it is obviously required that $\epsilon_0$ and $\epsilon_1$ must remain finite. However, it is always possible to choose $\gamma$ to set $\epsilon_1=0$. The metric is then 
 \begin{equation}
 \d s^2= {\gamma^2+p^2\over(1-\alpha\gamma p)^2} \Bigg[
 \tilde Q\,\d t^2 -{1\over\tilde Q}\,\d q^2 
  -{P\over(\gamma^2+p^2)^2} \Big(\d\psi+2\gamma q\,\d t \Big)^2 
  -{1\over P}\,\d p^2 \Bigg] \,, 
 \label{nonExpMetric}
 \end{equation} 
 where \ $\tilde Q= \epsilon_0-\epsilon_2q^2$ \ and 
 $$ P= k +2np -\epsilon p^2 +2\alpha mp^3
-\big[\alpha^2(k+e^2+g^2)+\Lambda/3\big]p^4 \,.  $$

It can be seen that the first terms in (\ref{nonExpMetric}) represent a (timelike) surface of constant curvature whose sign is that of $\epsilon_2$. It is then possible to use a linear transformation in $q$ (and $\psi$) together with a rescaling to set $\tilde Q$ to one of three distinct canonical forms 
 $$ 1-q^2, \qquad q^2-1, \qquad 1. \leqno{\qquad\tilde Q\,:} $$ 
 To achieve this, we require that 
 $$ \epsilon= -\epsilon_2 +6\alpha n\gamma 
 +2(3\alpha^2k+\Lambda)\gamma^2 \,.  $$ 
 In addition, $\gamma$ has been chosen to satisfy the equation 
 $$ 3m +(\epsilon_2-2\epsilon)\gamma +3\alpha n\gamma^2=0 \,, $$ 
 This equation relates $\gamma$ closely to the parameter $m$. It is therefore appropriate to generally make the relabeling $\gamma=\tilde m$. We also have the constraint that 
 $$ k+e^2+g^2 -\tilde mm +{\textstyle{1\over6}}(\epsilon+\epsilon_2)\tilde m^2 
 =\kappa^2\epsilon_0 \,, $$ 
 in the limit as $\kappa\to0$. This effectively determines~$k$. 
(In the alternative forms for $\tilde Q$ with different values of $\epsilon_0$ and $\epsilon_1$, the coordinate $q$ only covers a restricted range of the complete space-time. 
To transform between the canonical types and these others, a transformation of $\psi$ is also required, as in the similar situation described in detail in subsection~\ref{TNUTcases}.)

This family of solutions is thus characterised by one discrete parameter $\epsilon_2=1,-1,0$ and six continuous parameters $m$, $n$, $e$, $g$, $\Lambda$ and $\alpha$. Of course, the range of $p$ must be chosen such that $P>0$ with conformal infinity now at $p=(\alpha\tilde m)^{-1}$. But it is not now possible to make a linear transformation of $p$ to simplify the roots of $P$ without modifying the form of the metric (\ref{nonExpMetric}).

It can be seen that the conformal factor can be simplified by putting 
 \begin{equation} 
 p ={\tilde p-\alpha\tilde m^3\over1+\alpha\tilde m\tilde p} \,, 
 \label{transformp} 
 \end{equation}
 and applying a general rescaling. With this, $P$ remains a quartic but with coefficients which are complicated combinations of the above parameters.

This family of solutions was given by Carter \cite{Carter68} as his form
[$\tilde B(-)$]. It was also obtained by Pleba\'nski \cite{Pleb75} who referred to these as anti-NUT metrics since the transformation (\ref{trans1B}) is analogous to (\ref{trans1A}) which gives rise to generalised NUT metrics. For the vacuum case with $\Lambda=0$, these solutions are of Kinnersley's class~IV. 
An alternative form of the metric for this case was given by Kinnersley \cite{Kinnersley75}.

\subsubsection{The case when $\alpha=0$}

For the case in which $\alpha=0$ the metric (\ref{nonExpMetric}) becomes 
 \begin{equation}
 \d s^2= \rho^2\left(\tilde Q\,\d t^2 -{1\over\tilde Q}\,\d q^2 \right)
  -{P\over\rho^2} \Big(\d\psi+2\tilde m q\,\d t \Big)^2 
  -{\rho^2\over P}\,\d p^2 \,, 
 \label{nonAccnonExpMetric}
 \end{equation} 
  where \ $\rho^2= \tilde m^2+p^2$, \ 
$P= k +2np -\epsilon p^2 -{1\over3}\Lambda p^4$ \ in which \ $\epsilon=2\Lambda\tilde m^2-\epsilon_2$, \ and $\tilde m$ is chosen to satisfy the equation \ ${4\over3}\Lambda\tilde m^3 -\epsilon_2\tilde m -m =0$. \ 
It is also necessary that $k$ is given by \ 
$k= -e^2-g^2-\epsilon_2\tilde m^2 +\Lambda\tilde m^4 +\epsilon_0\kappa^2$, \ 
which is taken in the limit as $\kappa\to0$.

If $\tilde m\ne0$, these solutions are nonsingular. The function $\tilde P$ is a quartic if $\Lambda\ne0$ (otherwise it is at most a quadratic). It may therefore have up to four roots and the range of $p$ must be chosen such that $\tilde P>0$. The metric for these solutions is now given by (\ref{nonAccnonExpMetric}) in which $\tilde Q$ takes one of its canonical forms, $\rho^2 = p^2+\tilde m^2$, and 
 \begin{equation}
 P= -(e^2+g^2+\epsilon_2\tilde m^2 -\Lambda\tilde m^4) +2np 
+(\epsilon_2-2\Lambda\tilde m^2) p^2 
-{\textstyle{1\over3}}\Lambda p^4. 
 \label{Pnonexpcase}
 \end{equation}

When the parameter $\tilde m$, the charge parameters $e$ and $g$, and the cosmological constant all vanish (with $n\ne0$), these solutions are the $B$-metrics \cite{EhlersKundt62} in different coordinates: $BI$ when $\epsilon_2=1$, $BII$ when $\epsilon_2=-1$, and $BIII$ when $\epsilon_2=0$. When these additional parameters are non-zero, they represent a large family of distinct space-times. These depend, not only on the canonical form of $\tilde Q$, but also on the possible roots of $P$ and the range of $p$ that is chosen to ensure that $P(p)>0$.

For the case in which $\epsilon_2=1$ (i.e. when $\tilde Q=1-q^2$), we can apply the transformation 
 \begin{equation}
 p=r, \qquad q=\cos\theta, \qquad \psi=-\phi. 
 \end{equation} 
 This gives the line element 
 \begin{equation} 
 \d s^2= 
(r^2+\tilde m^2)\Big(\sin^2\theta\,\d t^2 -\d\theta^2 \Big)
-{P\over r^2+\tilde m^2}\Big(\d\phi
-2\tilde m\,\cos\theta\,\d t\Big)^2 -{r^2+\tilde m^2\over P}\,\d r^2 
 \label{BII} 
 \end{equation}  
 where 
 $$ P= -(e^2+g^2 +\tilde m^2 -\Lambda\tilde m^4) +2nr 
+(1-2\Lambda\tilde m^2)r^2 -{\textstyle{1\over3}}\Lambda r^4.  $$ 
 When $\tilde m=e=g=\Lambda=0$, $n\ne0$, this is the more familiar form of the $BI$-metric. However, it is not necessary or desirable to transform $q$ so that it is restricted to the range $q\in[-1,1]$. It is better to retain a general form which can be extended through the horizons at $q=1$ and $q=-1$.

Since the repeated principal null congruence is non-expanding and non-twisting (as well as being geodesic and shear-free), this family of solutions must belong to the type~D solutions of Kundt's class (see section \S31.7 of \cite{SKMHH03}). Indeed, the metric (\ref{nonAccnonExpMetric}) can be written in the standard form of the Kundt solutions using the transformation 
 $$ z=p, \qquad 
 y=\psi+2\tilde m\int{q\over\tilde Q}\,\d q, \qquad 
 u=t-\int{\d q\over\tilde Q},\qquad 
 v=\rho^2\,q, $$ 
 and putting 
 $$ \sqrt2\,\zeta=x+iy, \qquad x=\int{\rm P}^2(z)\d z, \qquad \hbox{where} 
 \qquad {\rm P}^2={\rho^2\over P}. $$ 
 This leads to the metric 
 \begin{equation} 
 \d s^2=2\d u\Big(\d v+W\,\d\zeta +\bar W\,\d\bar\zeta +H\,\d u\Big)
-{2\d\zeta\,\d\bar\zeta\over{\rm P}^2}, 
 \label{Kundtclass} 
 \end{equation}  
 where 
 $$ \begin{array}{l}
 {\displaystyle {\rm P}^2= {z^2+\tilde m^2\over 
-(e^2+g^2+\epsilon_2\tilde m^2 -\Lambda\tilde m^4) +2nz 
+(\epsilon_2-2\Lambda\tilde m^2)z^2 
-{1\over3}\Lambda z^4 } } \,, \\[12pt]
 {\displaystyle W=-{\sqrt2\,v\over(z+i\tilde m){\rm P}^2} } \,, \\[12pt]
 H={1\over2}\epsilon_0(z^2+\tilde m^2) 
-{\displaystyle \left[ {\epsilon_2\over2(z^2+\tilde m^2)} 
 +{2\tilde m^2\over(z^2+\tilde m^2)^2{\rm P}^2} \right]v^2 } .
 \end{array} $$ 
 This is a generalization of the solution given as equations (31.41) and (31.58)
in \cite{SKMHH03} which now includes a cosmological constant and an additional
discrete parameter. Further, there is a different identification of the
parameters $m$ and $n$.

\subsection{Non-expanding cases with $\omega=0$}
\label{nonexpomegazero}

Let us now consider the remaining exceptional case in which we again start with the metric (\ref{PleDemMetric}), but now with $\omega=0$: i.e. we start with the metric (\ref{twistfreeMetric}) with (\ref{twistfreePQeqns}). We then construct non-expanding solutions for this case by applying the transformation 
 $$ r=\gamma+\kappa q, \qquad \tau={\gamma^2\over\kappa}\,t, $$ 
 and taking the limit in which $\kappa=0$. This procedure leads to the line element 
  $$ \d s^2= {\gamma^2\over(1-\alpha\gamma p)^2}\left[ \tilde Q\,\d t^2 -{1\over\tilde Q}\,\d q^2 -P\,\d\sigma^2 -{1\over P}\,\d p^2 \right], $$ 
 where 
 $$ \begin{array}{l}
 P= k +2n'p -\epsilon p^2 +2\alpha mp^3 -\alpha^2(e^2+g^2)p^4 \,, \\[8pt] 
 \tilde Q= \epsilon_0+\epsilon_1q-\epsilon_2q^2 \,, 
 \end{array} $$  
 and 
 $$ \begin{array}{l}
 \epsilon_0= \kappa^{-2}(e^2+g^2 -2m\gamma +\epsilon\gamma^2 -2\alpha n'\gamma^3 -\alpha^2k\gamma^4
-{1\over3}\Lambda\gamma^4) \,, \\[6pt] 
 \epsilon_1= 2\kappa^{-1}(-m +\epsilon\gamma -3\alpha n'\gamma^2 -2\alpha^2k\gamma^3 -{2\over3}\Lambda\gamma^3) \,, \\[6pt] 
 \epsilon_2= -\epsilon +6\alpha n'\gamma +6\alpha^2k\gamma^2 
 +2\Lambda \gamma^2 \,. 
 \end{array} $$ 
 Thus, $\tilde Q$ is again an arbitrary quadratic which, for finite coefficients determines the appropriate choice of $\gamma$ and provides a constraint on the other parameters.

The case in which $\alpha=0$ will be dealt with in the following section. 
However, when $\alpha\ne0$, the conformal factor can be removed by the transformation 
 $$ 1-\alpha\gamma p=\gamma\,\tilde p^{-1}. \qquad 
 \sigma=\alpha\,\tilde\sigma \,. $$ 
 This takes the metric to the form 
 \begin{equation}
 \d s^2= \tilde p^2\left(\tilde Q\,\d t^2 -{1\over\tilde Q}\,\d q^2 \right)
  -{\tilde P\over\tilde p^2} \d\tilde\sigma^2 
  -{\tilde p^2\over\tilde P}\,\d\tilde p^2 ,
 \label{nonExpgamma0Metric}
 \end{equation} 
 where $\tilde Q$ may take one of the three standard canonical forms and 
$\tilde P(\tilde p)$ is a quartic function whose coefficients are different combinations of the parameters $m$, $n'$, $\epsilon$, $k$, $e$, $g$ and $\gamma$ to that given previously. This metric is in fact equivalent to (\ref{nonExpMetric}) after applying the transformation (\ref{transformp}) in the singular case in which $\gamma=0$ (i.e. $m=0$) in the notation of that section. However, the coefficients of the quartic $\tilde P$ are different.

This family of solutions has a curvature singularity at $\tilde p=0$. However, since $\tilde P$ is generally a quartic and we require that $\tilde P>0$, different ranges of $\tilde p$ are possible. The different possibilities which arise therefore represent various special cases of both singular and non-singular space-times.

\section{Non-twisting and non-accelerating solutions} 

\subsection{The metric (\ref{PleDemMetric}) with $\alpha=\omega=0$}

Let us finally consider the case when both the acceleration and the twist vanish.
Putting $\alpha=\omega=0$ accordingly, the metric (\ref{PleDemMetric}) becomes 
 \begin{equation}
 \d s^2= {Q\over r^2}\,\d\tau^2 -{r^2\over Q}\,\d r^2 
-r^2\left( {1\over P}\,\d p^2 +P\,\d\sigma^2 \right) \,,
 \label{ntwistnaccn}
 \end{equation} 
 where 
 \begin{equation}
 P =k +2n'p -\epsilon p^2, \qquad
 Q =(e^2+g^2) -2mr +\epsilon r^2 
-{\textstyle{1\over3}}\Lambda r^4, 
 \label{ntwistnaccnPQ}
 \end{equation}  
 as can also be seen from (\ref{twistfreeMetric}) and (\ref{twistfreePQeqns}). 
Apart from the possible presence of a non-zero cosmological constant and hence
curvature scalar, the only non-zero components of the curvature tensor take the
form 
 \begin{equation} 
 \Psi_2=-{m\over r^3} +{e^2+g^2\over r^4},
\qquad\qquad \Phi_{11}= {e^2+g^2\over2r^4}, 
 \label{ntwistnaccnCurv}
 \end{equation} 
 since the parameter $n$ must vanish in this case.

The surfaces on which $\tau$ and $r$ are constant have constant curvature $\epsilon$, so $P$ may be set to one of the standard three canonical forms. 
For the case in which $\epsilon=1$, $P=1-p^2$, and we can put 
 $$ p=\cos\theta, \qquad \sigma=-\phi, \qquad \tau=t, $$ 
 and the metric (\ref{ntwistnaccn}) then takes the familiar form of the
Reissner--Nordstr\"om--de~Sitter solution 
 $$ \begin{array}{l}
 \d s^2={\displaystyle \left(1-{2m\over r} +{e^2+g^2\over r^2}
+{\Lambda\over3}r^2\right)\!\d t^2- \left(1-{2m\over r} +{e^2+g^2\over r^2}
+{\Lambda\over3}r^2\right)^{-1}\!\d r^2 } \\[12pt]
 \hskip15pc -r^2(\d\theta^2+\sin^2\theta\,\d\phi^2).
 \end{array} $$
 This obviously includes the Schwarzschild, Reissner--Nordstr\"om, de~Sitter and
anti-de~Sitter solutions and their various combinations. Of course, this is the particular nontwisting subcase of the metric discussed in section~\ref{KNNUTdS} (for $a=0=l$).

Associated, but less physically significant, cases occur when $\epsilon=0$ ($P=1$) and when $\epsilon=-1$ ($P=p^2-1$). 
For the vacuum case with $\Lambda=0$, the different permitted values
of $\epsilon$ cover the three versions of the so-called $A$-metrics~\cite{EhlersKundt62}.

\subsection{Extension to include non-expanding solutions} 
\label{confflat}

The above family of solutions covers all non-accelerating and non-twisting type~D solutions with non-zero expansion. Interestingly, it is also possible to perform a coordinate transformation of (\ref{ntwistnaccn}) which, in a certain limit, includes an associated family of non-expanding type~D and conformally flat solutions.

Starting with the metric (\ref{ntwistnaccn}) with (\ref{ntwistnaccnPQ}), let us
consider the coordinate transformation which involves shifts in both $r$ and $p$ simultaneously 
 \begin{equation}
 p=\beta+\kappa\tilde p, \qquad r=\gamma+\kappa\tilde q, \qquad
\sigma=\kappa^{-1}\tilde\sigma, \qquad \tau=b^2\kappa^{-1}\tilde\tau, 
 \label{trans1C} 
 \end{equation}  
 where $\beta$, $\gamma$, $\kappa$ and $b$ are arbitrary constants. With this, the
metric (\ref{ntwistnaccn}) becomes 
 \begin{equation}
 \d s^2= {b^4\tilde Q\over(\gamma+\kappa\tilde q)^2}\,\d\tilde\tau^2 
-(\gamma+\kappa\tilde q)^2\left( {1\over\tilde Q}\,\d\tilde q^2
 +{1\over\tilde P}\,\d\tilde p^2 +\tilde P\d\tilde\sigma^2 \right), 
 \label{RNdSBR}
 \end{equation} 
 where
 $$  \begin{array}{l}
 \tilde P =a_0 +a_1\tilde p +a_2\tilde p^2, \\[6pt]
 \tilde Q =b_0+b_1\tilde q+b_2\tilde q^2 -{4\over3}\Lambda
\gamma\kappa\tilde q^3
 -{1\over3}\Lambda\kappa^2\tilde q^4,
 \end{array} $$ 
 and 
 $$ \begin{array}{l}
 b_2= \epsilon -2\Lambda \gamma^2 \\[6pt] 
 b_1= 2\kappa^{-1}(-m +\epsilon\gamma
-{2\over3}\Lambda\gamma^3) \\[8pt] 
 b_0= \kappa^{-2}(e^2+g^2 -2m\gamma +\epsilon\gamma^2
-{1\over3}\Lambda\gamma^4) \\
 \end{array} \qquad
 \begin{array}{l}
 a_2=-\epsilon \\[6pt]
 a_1=2\kappa^{-1}(n'-\epsilon\beta) \\[6pt]
 a_0=\kappa^{-2}(k+2\beta n'-\epsilon\beta^2) \\
 \end{array} $$

As in the previous section, we proceed to the non-expanding case by setting $\kappa=0$. In this case, the metric becomes 
 \begin{equation}
 \d s^2= b^2\left( Y\,\d\tilde\tau^2 -{1\over Y}\,\d\tilde q^2 \right)
 -\gamma^2\left( \tilde P\,\d\tilde\sigma^2  +{1\over\tilde P}\,\d\tilde p^2
\right), 
 \label{productspaces}
 \end{equation} 
 where \ $Y=b^2\gamma^{-2}(b_0+b_1\tilde q+b_2\tilde q^2)$. \ Provided $b_2$ and $\epsilon$ are non-zero, the linear terms in $Y$ and $\tilde P$ can be removed by choosing the free parameters $\gamma$ and $\beta$ respectively such that 
 \begin{equation}
 {\textstyle{2\over3}}\Lambda\gamma^3 -\epsilon\gamma +m =0 \qquad \hbox{and} 
 \qquad \epsilon\beta-n'=0. 
 \label{gamma}
 \end{equation} 
 In addition, the constant terms in $Y$ and $\tilde P$ will only be bounded if the parameters of the solution satisfy the constraints 
 \begin{equation} 
 \begin{array}{l}
 e^2+g^2 -2m\gamma +\epsilon\gamma^2
-{1\over3}\Lambda\gamma^4= b_0\kappa^2, \\[6pt] 
 k+2\beta n'-\epsilon\beta^2=a_0\kappa^2, \\
 \end{array} 
 \label{constraints7} 
 \end{equation} 
 as $\kappa\to0$ with $b_0$ and $a_0$ bounded. Together with (\ref{gamma}), the second of these constraints determines the required value of $k$ for such solutions to occur, and the first constraint is a condition on the parameters $m$, $e$, $g$, $\epsilon$ and $\Lambda$.

In all cases, linear transformations in $\tilde p$ and $\tilde q$ can then be used, together with a rescaling, so that $\tilde P$ and $Y$ respectively each take one of the standard canonical forms. In this way, we obtain the family of solutions given by Carter \cite{Carter68} as his form [$D$], which has also been given by Pleba\'nski \cite{Pleb75} as his case~C.

From the above discussion, together with that given below, it can be seen that
the metric (\ref{RNdSBR}) explicitly contains both the family of
Schwarzschild--Reissner--Nord\-str\"om--de~Sitter space-times (when $\kappa\ne0$) and the space-times that are a product of two 2-dimensional spaces of constant curvature such as Bertotti--Robinson (when $\kappa=0$). For the case in which the cosmological constant vanishes, the solution (\ref{RNdSBR}) includes that of Ray and Wei \cite{RayWei77} (rediscovered by Halilsoy \cite{Halil93}) which combines the Schwarzschild, the
Reissner--Nordstr\"om and the Bertotti--Robinson solutions in a single metric.

\subsubsection{The Bertotti--Robinson solution}

Let us now investigate the metric (\ref{productspaces}) for the case in which $\Lambda=0$ and $g=0$. In particular, for the case in which $\epsilon>0$, we use a rescaling to put $\epsilon=1$. We may also choose \ $b=\gamma$ \ and then choose \ $\beta=n'$ \ and \ $\gamma=m$ \ to satisfy (\ref{gamma}). We then consider the case in which the parameters satisfy the constraints (\ref{constraints7}) with \ $k=-{n'}^2+\kappa^2$ in the limit as \ $\kappa\to0$. \ In this case \ $\tilde P=1-\tilde p^2$ \ and \ $Y=\tilde q^2$. \ In addition, we have \ $e^2=m^2$ \ so that we are now considering a degenerate limit of the extreme Reissner--Nordstr\"om solution. In this case, the Weyl tensor component $\Psi_2$ vanishes identically. The resulting space-time is a conformally flat non-null electrovacuum solution, which must be the Bertotti--Robinson solution. 
Indeed, with the coordinate substitutions $\tilde p=\cos\theta$, 
$\tilde\sigma=-\phi$ and $\tilde q=1/\tilde r$, the line element for this case becomes
 \begin{equation} 
 \d s^2={{e^2}\over{\tilde r^2}}(\d\tilde\tau^2-\d\tilde r^2-\tilde r^2\d\theta^2
-\tilde r^2\sin^2\theta\,\d\phi^2), 
 \label{BR1} 
 \end{equation}  
 which is the most familiar form of the Bertotti--Robinson solution.

\subsubsection{Other direct product space-times} 

For the remaining cases that are included in the metric (\ref{productspaces}) with the conditions (\ref{gamma}) and (\ref{constraints7}), we note that $\tilde P$ can always be put into one of its three distinct canonical forms which correspond to the different parametrizations of 2-spaces of constant curvature. 
In addition, with $\gamma$ chosen to satisfy (\ref{gamma}), $Y$ is given by 
 $$ Y =\epsilon_0 -\epsilon_2\tilde q^2, $$ 
 where $\epsilon_2$ is given by 
 $$ \epsilon\gamma^{-2} +\epsilon_2b^{-2}=2\Lambda, $$
 and $\epsilon_0$ satisfies the constraint (\ref{constraints7}) which becomes 
 \begin{equation} 
 \gamma^{-2}(e^2+g^2)-\epsilon+\Lambda\gamma^2 =\kappa^2\epsilon_0b^{-2}=0. 
 \label{condition9}
 \end{equation} 
 In fact, it is possible to choose $b$ so that $\epsilon_2$ is equal to $+1$, 
$-1$ or 0. Then, since $\gamma$ is already defined by (\ref{gamma}) in terms of $m$, $\epsilon$ and $\Lambda$, (\ref{condition9}) represents a constraint on the parameters that is required to avoid a divergence in the metric.

The non-zero components of the curvature tensor (apart from the Ricci scalar
which is proportional to the cosmological constant) are now 
 $$ \Psi_2=-{\textstyle{1\over3}}\Lambda, \qquad 
\Phi_{11}={\textstyle{1\over2}}\gamma^{-4}(e^2+g^2). $$ 
 (The terms containing $m$ and $e^2+g^2$ that occurred in $\Psi_2$ in (\ref{ntwistnaccnCurv}) have cancelled out in the limit $\kappa=0$, apart from a term proportional to the cosmological constant, through the choice and constraint (\ref{gamma}) and (\ref{condition9}).)

It can thus be seen that the metric (\ref{productspaces}) is the direct product of two 2-dimensional spaces of constant curvatures $\epsilon$ and $\epsilon_2$ of signatures $(+,-)$ and $(-,-)$ respectively. The values of these two parameters determine entirely the geometry of the space-time. As described e.g. by Ortaggio and
Podolsk\'y \cite{OrtPod02}, these represent the complete family of
Bertotti--Robinson, Narai and Pleba\'nski--Hacyan space-times. Excluding cases
with negative energy density, it is found that there are six possible geometries. 

Allowing for the possibilities of both electric and magnetic charges, this family of space-times essentially depends on only three free dynamical parameters, which may be taken as $e$, $g$ and~$\Lambda$. (The parameter $m$, which is necessarily non-zero, may be reintroduced through the definition of $\gamma$ and the constraint (\ref{condition9}).)

\section{Conclusions} 

We have analysed in detail the complete family of Pleba\'nski--Demia\'nski metrics and those that can be derived from them by taking limits after performing some specific coordinate transformations. These non-null solutions of the Einstein--Maxwell equations with a possible cosmological constant include all type~D solutions in which the principal null directions of the electromagnetic field (if included) are aligned with the repeated principal null directions of the Weyl tensor and the group orbits of the Killing vectors are non-null.

\begin{figure}[hbt]
\begin{center} \includegraphics[scale=0.9, trim=5 -5 5 -5]{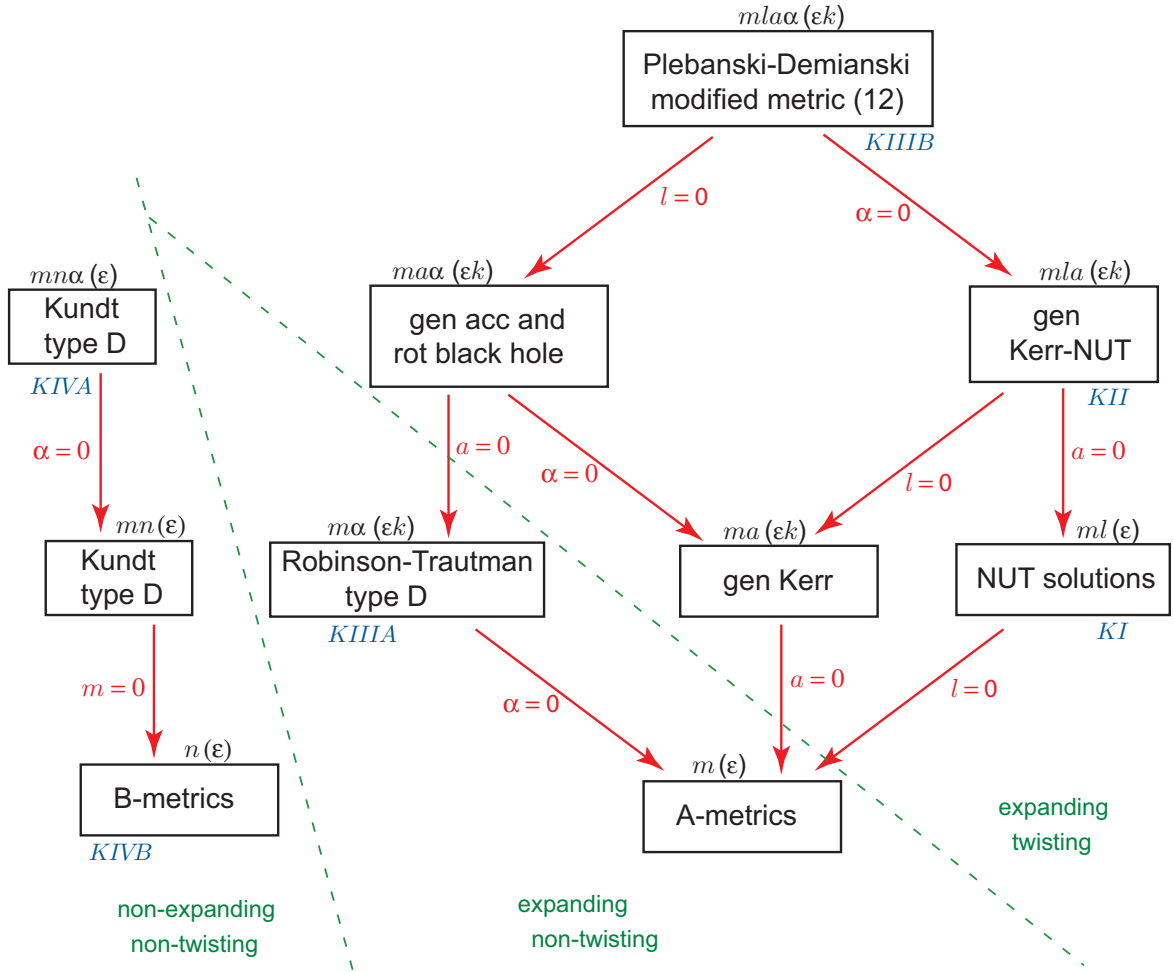}
\caption{\small Vacuum type~D solutions with $\Lambda=0$ and $m\ne0$. In the expanding case, these solutions generally have four continuous parameters $m$, $l$, $\alpha$ and $a$, and two auxiliary parameters $\epsilon$ and $k$ that can be scaled to any convenient specific values. The parameters associated with each solution are shown above each box -- those in brackets have discrete values. The abbreviation ``gen'' indicates that the named solution is generalised to include the associated solutions with other values of the discrete parameters. The Kinnersley class of each sub-family is also indicated below each box. All particular case have obvious generalizations which include charges and/or a cosmological constant. }
\label{vacuumPD}
\end{center}
\end{figure}

In general, we have found it convenient to treat separately the cases in which the repeated principal null congruences are either expanding or non-expanding. For the expanding case, the metric which includes all possibilities is given by (\ref{altPlebMetric}). For the non-expanding case, the general metric is given by (\ref{nonExpMetric}) although, when $\Lambda\ne0$, other type~D solutions are given by (\ref{productspaces}), which also includes certain associated conformally flat solutions. The non-expanding solutions can alternatively be expressed in the standard form for solutions of Kundt's class, namely~(\ref{Kundtclass}).

The structure of the complete family of Pleba\'nski--Demia\'nski solutions for the vacuum case with $e=g=\Lambda=0$ is illustrated in figure~\ref{vacuumPD}. For this case, it seems to be most convenient to represent the expanding solutions in terms of four continuous parameters $m$, $l$, $\alpha$ and $a$, and two auxiliary parameters $\epsilon$ and $k$ that can be scaled to any convenient specific values. For the case in which the line element has a component in which the coordinates span surfaces of positive curvature, the classification of these solutions has been illustrated in more detail in figure~\ref{AccRotBHstructure}. 
In that case, the family of Robinson--Trautman type~D solutions is represented by the $C$-metric. It was appropriate there to distinguish the cases in which the Kerr-like rotation parameter is greater or less than the NUT parameter, as these cases have significantly different singularity structures. However, this distinction is not continued in figure~\ref{vacuumPD}. 
Here, the solutions are generalized to include those with surfaces of alternative curvatures and the non-expanding solutions are also included. All the particular cases illustrated in figure~\ref{vacuumPD} have obvious generalizations with non-zero charge parameters and/or a non-zero cosmological constant.

In general, these solutions are characterized by two generally quartic functions whose coefficients are related to the physical parameters of the space-time. 
However, although it is traditional to use an available coordinate freedom to remove the linear term in one of these quartics, it has been found to be much more helpful to use this freedom to simplify the roots of at least one of these functions. This significantly simplifies the calculations involved in interpreting these solutions and in identifying the physical interpretation of the coefficients. In particular, as also shown in \cite{GriPod05}, it is found that the Pleba\'nski--Demia\'nski parameter $n$ is not the NUT parameter, but is related to it by (\ref{n}).

Of course, most of the solutions in this family do not yet have any known physical significance as models of the gravitational fields of realistic sources. Nevertheless, they provide a most important family of model space-times that can at least be interpreted geometrically.

\section*{Acknowledgements}

This work was supported in part by a grant from the EPSRC.


\begin{thebibliography}{99}


\bibitem{PleDem76} Pleba\'nski, J. F. and Demia\'nski, M. (1976). Rotating
charged and uniformly accelerating mass in general relativity, {\sl Ann. Phys.
(NY)}, {\bf 98}, 98--127. 

\bibitem{Kinnersley69} Kinnersley, W. (1969). Type D vacuum metrics, {\sl J.
Math. Phys.}, {\bf 10}, 1195--1203. 

\bibitem{WeirKerr77} Weir, G. J. and Kerr, R. P. (1977). Diverging type-D
metrics, {\sl Proc. Roy. Soc. A}, {\bf 355}, 31--52. 

\bibitem{DebKam80} Debever, R. and Kamran, N. (1980). Coordonn\'ees
sym\'etriques et coordonn\'ees isotropes des solutions de type D des
\'equations d'Einstein--Maxwell avec constante cosmologique, {\sl Bull. Cl.
Sci. Acad. Roy. Belg.}, {\bf 66}, 585--599. 

\bibitem{IshMiy82} Ishikawa, K. and Miyashita, T. (1982). Classification of
Petrov type D empty Einstein spaces with diverging null geodesic congruences,
{\sl Prog. Theor. Phys.}, {\bf 67}, 828--843. 

\bibitem{HawRoss95} Hawking, S. W. and Ross, S. F. (1995). Pair production of
black holes on cosmic strings, {\sl Phys. Rev. Lett.}, {\bf 75}, 3382--3385. 

\bibitem{ManRos95} Mann, R. B. and Ross, S. F. (1995). Cosmological
production of charged black hole pairs, {\sl Phys. Rev. D} {\bf 52},
2254--2265. 

\bibitem{BooMan99} Booth, I. S. and Mann, R. B. (1999). Cosmological pair production of charged and rotating black holes, {\sl Nucl. Phys. B}, {\bf 539}, 267--306. 

\bibitem{Carter68} Carter, B. (1968). Hamilton--Jacobi and Schr\"odinger
separable solutions of Einstein's equations, {\sl Commun. Math. Phys.} {\bf
10}, 280--310. 

\bibitem{DeKaMcL84} Debever, R., Kamran, N. and McLenaghan, R. G. (1984).
Exhaustive integration and a single expression for the general solution of the
type D vacuum and electrovac field equations with cosmological constant for a
nonsingular aligned Maxwell field, {\sl J. Math. Phys.}, {\bf 25}, 1955--1972. 

\bibitem{GarciaD84} Garc\'{\i}a D., A. (1984). Electrovac type D solutions
with cosmological constant, {\sl J. Math. Phys.} {\bf 25}, 1951--1954. 

\bibitem{HongTeo03} Hong, K. and Teo, E. (2003). A new form of the
$C$-metric, {\sl Class. Quantum Grav.},
{\bf 20}, 3269--3277. 

\bibitem{HongTeo05} Hong, K. and Teo, E. (2005). A new form of the rotating
$C$-metric, {\sl Class. Quantum Grav.}, {\bf 22}, 109--117. 

\bibitem{KowPleb77} Kowalczy\'nski, J. K. and Pleba\'nski, J. F. (1977).
Metric and Ricci tensors for a certain class of space-times of type D, {\sl
Int. J. Theoret. Phys.}, {\bf 16}, 371--388; erratum {\bf 17}, 388 (1978). 

\bibitem{SKMHH03} Stephani, H., Kramer, D., MacCallum, M., Hoenselaers, C. and
Herlt, E. (2003)  {\sl Exact solutions of Einstein's field equations, 2nd
edition}, (Cambridge University Press). 

\bibitem{Debever71} Debever, R. (1971). On type D expanding solutions of
Einstein--Maxwell equations, {\sl Bull. Soc. Math. Belg.} {\bf 23}, 360--76. 

\bibitem{PodGri01} Podolsk\'y, J. and Griffiths, J. B. (2001). Uniformly
accelerating black holes in a de~Sitter universe, {\sl Phys. Rev. D}, {\bf
63}, 024006. 

\bibitem{GarPle82} Garc\'{\i}a D\'{\i}az, A. and Pleba\'nski, J. F. (1982).
Solutions of type D possessing a group with null orbits as contractions of
the seven-parameter solution, {\sl J. Math. Phys.}, {\bf 23}, 1463--1465. 

\bibitem{Leroy78} Leroy, J. (1978). Sur une classe d'espaces-temps solutions des equations d'Einstein--Maxwell. 
{\sl Bull. Acad. Roy. Belg. Cl. Sci.}, {\bf 64}, 130. 

\bibitem{GarSal83} Garc\'{\i}a D\'{\i}az, A. and Salazar, I. H. (1983).
All null orbit type d electrovac solutions with cosmological constant, {\sl J.
Math. Phys.}, {\bf 24}, 2498--2503. 

\bibitem{DeKaMcL83} Debever, R., Kamran, N. and McLenaghan, R. G. (1983). A
single expression for the general solution of Einstein's vacuum and
electrovac field equations with cosmological constant for Petrov type~D
admitting a non-singular aligned Maxwell field, {\sl Phys. Lett. A}, {\bf 93},
399--402. 

\bibitem{GarMac98} Garc\'{\i}a D., A. and Macias, A. (1998). Black holes as
exact solutions of the Einstein--Maxwell equations of Petrov type D, in {\sl
Black holes: Theory and observation}, Lecture notes in Physics {\bf 514}, eds
F. W. Hehl, C. Kiefer and R. J. K. Metzler, (Springer) p205. 

\bibitem{Pleb75} Pleba\'nski, J. F. (1975). A class of solutions of the
Einstein--Maxwell equations, {\sl Ann. Phys. (NY)}, {\bf 90}, 196--255. 

\bibitem{GriPod05} Griffiths, J. B. and Podolsk\'y, J. (2005).  Accelerating and rotating black holes, {\sl Class. Quantum Grav.}, {\bf 22}, 3467--3479. 

\bibitem{NewTamUnt63} Newman, E. T., Tamburino, L. A. and Unti, T. (1963).
Empty-space generalization of the Schwarzschild metric, {\sl J. Math, Phys.},
{\bf 4}, 915--923. 

\bibitem{Bonnor69} Bonnor, W. B. (1969). A new interpretation of the NUT
metric in general relativity, {\sl Proc. Camb. Phil. Soc.}, {\bf 66},
145--151. 

\bibitem{FarZim80b} Farhoosh, H. and Zimmerman, R.L. (1980). Surfaces of
infinite red-shift around a uniformly accelerating and rotating particle,
{\sl Phys. Rev. D} {\bf 21}, 2064--2074. 

\bibitem{BicPra99} Bi\v{c}\'ak, J. and Pravda, V. (1999). Spinning $C$
metric: Radiative spacetime with accelerating, rotating black holes, {\sl
Phys. Rev. D}, {\bf 60}, 044004. 

\bibitem{Pravdas02} Pravda, V. and Pravdov\'a, A. (2002). On the spinning
C-metric.  In {\sl Gravitation: Following the Prague Inspiration}
  (A celebration of the 60th Birthday of J. Bi\v c\'ak), 
 eds O. Semer\'ak, J. Podolsk\'y and M. \v{Z}ofka,
  (World Scientific), pp 247--262. 

\bibitem{PodGri06} Podolsk\'y, J. and Griffiths, J. B. (2006). Accelerating Kerr--Newman black holes in (anti-)de~Sitter space-time. In preparation. 

\bibitem{HawEll73} Hawking, S. W. and Ellis, G. R. F. (1973). {\sl The large
scale structure of space-time}. Cambridge University Press. 

\bibitem{GriPod06} Griffiths, J. B. and Podolsk\'y, J. (2006). Global aspects of accelerating and rotating black hole space-times. {\sl Submitted to Class. Quantum Grav.} 

\bibitem{DiaLem03a} Dias, \'O. J. C. and Lemos, J. P. S. (2003). Pair of
accelerated black holes in an anti-de~Sitter background: the AdS $C$ metric,
{\sl Phys. Rev. D}, {\bf 67}, 064001, 1--19.   

\bibitem{DiaLem03b} Dias, \'O. J. C. and Lemos, J. P. S. (2003). Pair of
accelerated black holes in a de~Sitter background: the dS $C$ metric,
{\sl Phys. Rev. D}, {\bf 67}, 084018. 

\bibitem{KrtPod03} Krtou\v{s}, P. and Podolsk\'y, J. (2003). Radiation from
accelerating black holes in a de Sitter universe, {\sl Phys. Rev. D}, {\bf 68},
024005. 

\bibitem{PoOrKr03} Podolsk\'y, J, Ortaggio, M. and  Krtou\v s, P. (2003).
  Radiation from accelerated black holes in an anti-de Sitter universe.
  {\sl Phys. Rev. D}, {\bf 68}, 124004. 

\bibitem{Carter70} Carter, B. (1970). The commutation property of a
stationary, axisymmetric system, {\sl Commun. Math. Phys.} {\bf 17},
233--238. 

\bibitem{GibHaw77} Gibbons, G. W. and Hawking, S. W. (1977). Cosmological
event horizons, thermodynamics, and particle creation, {\sl Phys. Rev. D},
{\bf 15}, 2738--2751. 

\bibitem{Ruban72} Ruban, V. A. (1972). Nonsingular
Taub--Newman--Unti--Tamburino metrics. {\sl Dokl. Acad. Nauk CCCP}, {\bf
204}, 1085--1089: {\sl Sov. Phys. Dokl.}, {\bf 17}, 568--571. 

\bibitem{EhlersKundt62} Ehlers, J. and Kundt, W. (1962). Exact solutions of the
gravitational field equations. In {\sl Gravitation: an introduction to
current research}, (ed. L. Witten). Wiley, New York. pp 49--101. 

\bibitem{Kinnersley75} Kinnersley, W. (1975). Recent progress in exact solutions. In {\sl General relativity and gravitation}, Proc. GR7. eds G. Shaviv and J. Rosen, (Wiley), pp. 109--35. 

\bibitem{RayWei77} Ray, J. R. and Wei, M. S. (1977). A solution-generating
theorem with applications in general relativity, 
 {\sl Nuovo Cimento B}, {\bf 42}, 151--164. 

\bibitem{Halil93} Halilsoy, M. (1993). Interpolation of the Schwarzschild and
Bertotti--Robinson solutions, {\sl Gen. Rel. Grav.}, {\bf 25}, 275--280. 

\bibitem{OrtPod02} Ortaggio, M and Podolsk\'y, J. (2002).
  Impulsive waves in electrovac direct product spacetimes with $\Lambda$.
  {\sl Class. Quantum Grav.}, {\bf 19}, 5221--5227. 


\end{thebibliography}
\end{document}